\documentclass{article}

\usepackage{PRIMEarxiv}

\usepackage[utf8]{inputenc} 
\usepackage[T1]{fontenc}    
\usepackage{hyperref}       
\usepackage{url}            
\usepackage{booktabs}       
\usepackage{amsfonts}       
\usepackage{nicefrac}       
\usepackage{microtype}      
\usepackage{graphicx}       
\usepackage{amsmath}
\graphicspath{{media/}{./}}  
\usepackage{pgfplots}
\usepackage{xcolor}
\usepackage{subcaption}
\pgfplotsset{compat=1.18}
\usetikzlibrary{shapes.geometric, arrows.meta, positioning, calc}

\definecolor{light-green}{HTML}{e0f0d8}
\definecolor{light-blue}{HTML}{d9ecfa}
\definecolor{light-orange}{HTML}{ffeacc}
\definecolor{light-pink}{HTML}{fad4dd}
\definecolor{light-purple}{HTML}{e5c0ff}
\definecolor{light-yellow}{HTML}{fcfad9}
\definecolor{light-red}{HTML}{fcd1c0}
\definecolor{light-teal}{HTML}{b2e0e0}
\definecolor{light-coral}{HTML}{f5b7a3}
\definecolor{orig-green}{HTML}{2ecc71}
\definecolor{variant-red}{HTML}{e74c3c}
\definecolor{acad-blue}{HTML}{3498db}
\definecolor{industry-orange}{HTML}{e67e22}
\definecolor{pre-blue}{HTML}{4C72B0}
\definecolor{post-orange}{HTML}{DD8452}

\title{TRACING THE EVOLUTION OF WORD EMBEDDING TECHNIQUES IN NATURAL LANGUAGE PROCESSING}

\author{
  Minh Anh Nguyen \\
  Texas Tech University \\
  Lubbock, Texas \\
  \texttt{MinhAnh.Nguyen@ttu.edu} \\
   \And
  Kuheli Sai \\
  University of Pittsburgh \\
  Pittsburgh, Pennsylvania \\
  \texttt{kuheli.sai@pitt.edu} \\
     \And
  Minh Nguyen \\
  Florida Atlantic University \\
  Boca Raton, Florida \\
  \texttt{minhnguyen@fau.edu} \\
}

\begin{document}
\maketitle

\begin{abstract}
This work traces the evolution of word-embedding techniques within the natural language processing (NLP) literature. We collect and analyze 149 research articles spanning the period from 1954 to 2025, providing both a comprehensive methodological review and a data-driven bibliometric analysis of how representation learning has developed over seven decades. Our study covers four major embedding paradigms, statistical representation-based methods (one-hot encoding, bag-of-words, TF-IDF), static word embeddings (Word2Vec, GloVe, FastText), contextual word embeddings (ELMo, BERT, GPT), and sentence/document embeddings, critically discussing the strengths, limitations, and intellectual lineage connecting each category. Beyond the methodological survey, we conduct a formal era comparison using GPT-3's release as a dividing line, applying seven hypothesis tests to quantify shifts in research focus, collaboration patterns, and institutional involvement. Our analysis reveals a dramatic post-GPT-3 paradigm shift: contextual and sentence-level methods now dominate at $6.4\times$ the odds of the pre-GPT-3 era, mean team sizes have grown significantly ($p = 0.018$), and 30 entirely new techniques have emerged while 54 pre-GPT-3 methods received no further attention. These findings, combined with evidence of rising industry involvement, provide a quantitative account of how the field's epistemic priorities have been reshaped by the advent of large language models.
\end{abstract}


\section{Introduction}

Despite the importance of word embeddings in modern natural language processing (NLP), the historical development of these techniques has not been systematically documented in a way that connects early statistical representations to today's contextual and generative models. Prior survey studies often focus on particular model families, such as prediction-based versus count-based embeddings, or neural embeddings alone, without situating them within the broader, decades-long evolution of representation learning. As a result, the research community lacks a unified narrative that explains how conceptual advances accumulated over time, how methodological trade-offs shifted across eras, and how each generation of embedding techniques addressed shortcomings of its predecessors. Our study fills this gap by synthesizing seven decades of model development and contextualizing individual innovations within a comprehensive historical trajectory.

A key premise of our work is the conceptual continuity that unites seemingly distinct embedding paradigms. Methods as early as TF–IDF or latent semantic analysis (LSA), though primitive by today's standards, encode many of the same principles that underlie deep contextual models, namely distributional semantics, dimensionality reduction, and the abstraction of linguistic information into geometric structures. Understanding this continuity not only demystifies the "jump" from static to contextual embeddings but also helps researchers appreciate why certain ideas persist and resurface in new forms. In doing so, our study not only catalogs the field's major contributions but also articulates the intellectual lineage connecting them.

To support this historical synthesis, our work contributes a uniquely data-driven analysis of the word embedding literature. By assembling a curated set of 149 papers spanning from 1954 to 2025, we not only identify major methodological milestones but also characterize trends in publication patterns, collaboration dynamics, and institutional influence. This broader lens reveals how the research landscape has transformed over time, for example, how early theoretical contributions from linguistics and information retrieval eventually gave way to neural approaches, and how institutional involvement gradually shifted from academic labs to industry-led research groups. These patterns, which remain largely undocumented in prior surveys, shed light on the socio-technical forces shaping the evolution of NLP methodologies.

Finally, a central motivation for this work lies in understanding how the emergence of large language models (LLMs) has redefined the role of word embeddings within NLP. Since the release of transformer-based systems such as BERT, GPT-3, and ChatGPT, embeddings have moved from being standalone components to functioning as internal representations embedded within massive, end-to-end models. This paradigm shift raises important questions: To what extent do traditional embedding methods remain relevant? Which foundational ideas continue to influence modern architectures? And how has the introduction of LLMs reshaped research attention, model design, and assumptions about semantic representation? By explicitly comparing pre-LLM and post-LLM eras, our analysis provides clarity on how the field's epistemic priorities have evolved.

In summary, the main contributions of this work are:
\begin{enumerate}
    \item A comprehensive historical survey of word embedding techniques spanning seven decades (1954--2025), organized into four paradigm categories and tracing the intellectual lineage from early statistical methods to modern contextual models.
    \item A curated dataset of 149 papers, each annotated along multiple dimensions (embedding category, originality, author affiliations, industry involvement, publication venue), enabling systematic quantitative analysis.
    \item A formal era comparison using GPT-3's release as a dividing line, with seven hypothesis tests that quantify statistically significant shifts in methodological focus, team composition, and collaboration patterns.
    \item A bibliometric analysis of publication trends, institutional contributions, geographic concentration, and the evolving role of industry in word embedding research.
\end{enumerate}

The remainder of this paper is organized as follows. Section~2 reviews prior survey literature and positions our contribution. Section~3 describes our data collection, search strategy, and annotation methodology. Section~4 presents the results, including descriptive statistics, detailed analyses of representative techniques from each category, and the pre- vs.\ post-GPT-3 era comparison. Section~5 discusses implications for researchers, societal impact, and ethical considerations. Section~6 concludes with a summary of findings, limitations, and directions for future work.

\section{Literature Review}

Since this paper is itself a survey of word embedding techniques, the individual methods and architectures (e.g., Word2Vec, GloVe, BERT) are reviewed in detail within the body of this work rather than in this section. Here, we instead situate our contribution relative to existing survey literature, identifying the scope, strengths, and limitations of prior reviews.

The study of word embeddings has attracted considerable scholarly attention, producing a rich body of survey literature that has evolved alongside the techniques themselves. Early efforts to systematize this field focused primarily on distinguishing between prediction-based and count-based paradigms. Almeida and Xex\'{e}o \cite{almeida2019word}, for instance, provide a foundational taxonomy centered on dense, distributed representations derived from the distributional hypothesis, reviewing approximately 20 major models including Word2Vec \cite{mikolov2013efficient} and GloVe \cite{pennington2014glove}. Around the same period, Wang et al. \cite{wang2020survey} extended this scope by incorporating deep learning perspectives, tracing the progression from static embeddings toward early contextual representations such as ELMo \cite{peters2018deep} and BERT \cite{devlin2019bert}. While these surveys offer valuable introductions to embedding architectures, they remain largely descriptive and do not incorporate quantitative bibliometric analysis or capture broader trends in the research landscape.

As the field matured, subsequent reviews began to address the transition from static to contextual embeddings more explicitly. Sezerer and Tekir \cite{sezerer2021survey} ground their survey in both distributional semantics and language modeling traditions, covering extensions such as sense embeddings and morphological embeddings. However, writing from the vantage point of 2021, their work necessarily omits later breakthroughs including GPT-3 \cite{brown2020language}, GPT-4 \cite{openai2023gpt4}, LLaMA \cite{touvron2023llama}, and multimodal models such as CLIP \cite{radford2021learning} and Flamingo \cite{alayrac2022flamingo}. This temporal limitation is inherent to any survey but underscores the need for updated syntheses that integrate the most recent advances.

Beyond these broad methodological surveys, several studies have examined embeddings through more specialized lenses. Rodriguez and Spirling \cite{rodriguez2022word} focus narrowly on hyperparameter choices for GloVe and Word2Vec within political science applications, offering domain-specific insights but limited generalizability. Similarly, Bollegala and O'Neill \cite{bollegala2022survey} concentrate on the niche area of meta-embedding learning, which combines multiple embedding sources to capture broader representations. While valuable for their respective audiences, these specialized perspectives do not attempt to provide a unified, cross-domain view of the embedding literature.

More recent surveys have endeavored to be comprehensive, yet gaps remain. Patil et al. \cite{patil2023survey} review a wide range of text representation methods in depth but focus almost exclusively on methodological aspects, leaving broader dimensions such as publication trends, institutional contributions, and collaboration patterns unexplored. Mahajan et al. \cite{mahajan2024revisiting} take a timely step by examining embeddings in the era of large language models and providing empirical comparisons; however, their scope is restricted to LLM-based techniques, omitting the historical trajectory that connects early statistical methods to contemporary contextual models.

Collectively, these prior surveys reveal a fragmented landscape: some emphasize architectural distinctions without historical context, others adopt domain-specific or technique-specific foci, and none combine methodological depth with data-driven bibliometric analysis. Our work addresses these gaps by synthesizing seven decades of embedding research, from early statistical representations through modern transformer-based models, while simultaneously providing quantitative insights into publication patterns, collaboration structures, and the shifting role of embeddings in the LLM era.

\section{Data Collection}

\subsection{Search Strategy}
We began the data collection process by first defining the scope of our dataset, restricting it to works that introduced new embedding methods or proposed enhancements or variants of existing ones, rather than applied studies that used embeddings in downstream domains. The objective of this stage was to assemble a solid dataset that ensured coverage across different categories of embeddings. To mitigate bias, our collection process combined automated search strategies with manual verification. Specifically, we relied on LLMs, academic databases, and citation tracking from recent literature review studies. In addition to identifying new papers, LLMs were also employed to trace the origins of specific word embedding methods by locating their original publications. This multi-pronged approach enabled us to cover both foundational contributions and the latest advancements in the field.

\noindent
\textbf{Data Sources and Iterative Discovery.} First, we utilized LLMs including ChatGPT, Gemini, and Claude to identify seminal works introducing novel techniques as well as recent methodological developments. To ensure accuracy and relevance, all suggested works were verified through Google Scholar before being added to a preliminary corpus. This corpus then served as a foundation for further exploration through citation tracking and complementary search strategies.

\noindent
Second, we systematically queried Google Scholar to retrieve literature on word embeddings across different time periods and categories. By experimenting with keyword variations (e.g., ``distributional semantics,'' ``contextual embeddings,'' ``pretrained language models''), we captured a broader range of works, including smaller-scale studies that provided valuable extensions and refinements of the original methods.

\noindent
Third, we examined recent review papers on word embeddings to identify overlooked studies. We analyzed reference lists from key survey papers published within the last five years, cross-referencing their cited works against our existing corpus. This backward citation tracking revealed foundational papers introducing lesser-known embedding techniques and recent innovations emerging between major reviews.

Finally, we conducted forward citation tracking from seminal works such as Word2Vec \cite{mikolov2013efficient}, GloVe \cite{pennington2014glove}, and BERT \cite{devlin2019bert} to capture derivative methods and improvements. Each newly identified paper underwent relevance assessment to ensure it met our inclusion criteria of introducing novel embedding methods or significant enhancements, rather than merely applying embeddings to specific domains.

\paragraph{Final Corpus Construction.} The final dataset comprises 149 papers, spanning the period from 1954 (Harris's foundational work on distributional structure) to 2025, after excluding applied studies focused solely on downstream domains. This collection captures both the earliest statistical approaches to word representation and the most recent advances in contextual embeddings. 

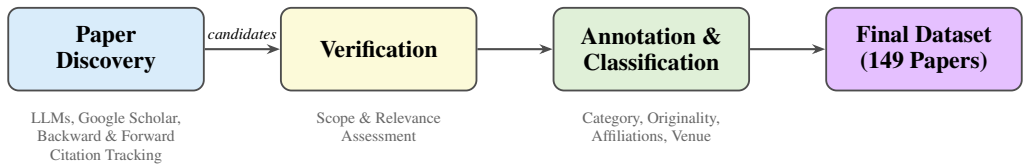
\begin{figure}[h!]
    \centering
    \begin{tikzpicture}[
        stage/.style={rectangle, rounded corners=4pt, draw=black!70, line width=0.7pt, minimum width=2.6cm, minimum height=1.1cm, align=center, font=\footnotesize\bfseries},
        arrow/.style={-{Stealth[length=5pt]}, thick, black!70},
        lbl/.style={font=\tiny\itshape, text=black, align=center},
    ]
    \node[stage, fill=light-blue] (discovery) {Paper\\Discovery};
    \node[stage, fill=light-yellow, right=1.0cm of discovery] (verify) {Verification};
    \node[stage, fill=light-green, right=1.0cm of verify] (annotate) {Annotation \&\\Classification};
    \node[stage, fill=light-purple, right=1.0cm of annotate] (final) {Final Dataset\\(149 Papers)};
    
    \draw[arrow] (discovery.east) -- node[above, lbl] {candidates} (verify.west);
    \draw[arrow] (verify.east) -- (annotate.west);
    \draw[arrow] (annotate.east) -- (final.west);
    
    \node[below=0.15cm of discovery, font=\tiny, align=center, text=black!60] {LLMs, Google Scholar,\\Backward \& Forward\\Citation Tracking};
    \node[below=0.15cm of verify, font=\tiny, align=center, text=black!60] {Scope \& Relevance\\Assessment};
    \node[below=0.15cm of annotate, font=\tiny, align=center, text=black!60] {Category, Originality,\\Affiliations, Venue};
    
    \end{tikzpicture}
    \caption{Data collection and annotation pipeline.}
    \label{fig:data_pipeline}
\end{figure}

\subsection{Annotation and Classification}

Each paper in the dataset was manually annotated along several dimensions to enable systematic quantitative analysis. The annotation was performed by a single researcher, with all classifications verified against the original publications. While the use of LLMs assisted in initial paper discovery, every classification decision was validated by reading the paper's abstract, methodology, and contribution sections.

\paragraph{Embedding Category.}
Each paper was assigned to one of four mutually exclusive embedding categories based on the representation type and granularity of the proposed technique:

\begin{itemize}
    \item \textbf{Statistical Representations}: Methods that produce sparse, high-dimensional vectors through frequency-based or distributional counting (e.g., One-hot encoding, Bag-of-Words, TF-IDF \cite{sparck1972statistical}, LDA \cite{blei2003latent}, LSA/LSI \cite{deerwester1990indexing}, HAL \cite{lund1996producing}). These techniques do not involve neural network training and derive representations directly from corpus statistics.
    \item \textbf{Static Word Embeddings}: Methods that learn dense, fixed-dimensional vectors for each word type, where each word receives a single representation regardless of context (e.g., Word2Vec \cite{mikolov2013efficient}, GloVe \cite{pennington2014glove}, FastText \cite{bojanowski2017enriching}). The defining characteristic is that the same word always maps to the same vector.
    \item \textbf{Contextual Word Embeddings}: Methods that generate token-level representations conditioned on the surrounding context, producing different vectors for the same word in different sentences (e.g., ELMo \cite{peters2018deep}, BERT \cite{devlin2019bert}, GPT \cite{radford2018improving}, Transformer-based architectures \cite{vaswani2017attention}). These models typically employ attention mechanisms or recurrent layers that incorporate contextual information.
    \item \textbf{Sentence/Document Embedding}: Methods whose primary output is a fixed-length vector representing an entire sentence, paragraph, or document rather than individual tokens (e.g., Doc2Vec \cite{le2014distributed}, Sentence-BERT \cite{reimers2019sentence}, Universal Sentence Encoder \cite{cer2018universal}, Word Mover's Distance \cite{kusner2015from}). This category includes both dedicated sentence encoders and methods that aggregate word-level embeddings into sequence-level representations.
\end{itemize}

\paragraph{Original vs.\ Variant/Enhancement.}
Each paper was classified as either Original Research or Variant/Enhancement based on the nature of its methodological contribution. A paper was classified as original if it introduced a fundamentally new embedding architecture, training paradigm, or representational framework. A paper was classified as a variant or enhancement if it primarily built upon, modified, or extended an existing method---for instance, by applying knowledge distillation to BERT (DistilBERT \cite{sanh2019distilbert}), adding more training data (RoBERTa \cite{liu2019roberta}), or combining two existing techniques. We acknowledge that this distinction involves a degree of subjective judgment, as no research is entirely independent of prior work; however, the classification reflects whether the paper's primary contribution was a novel architecture or an incremental improvement to an identified predecessor.

\paragraph{Author Affiliations and Industry Involvement.}
For each author, we recorded their first-listed institutional affiliation as reported in the paper. When an author listed multiple affiliations (e.g., a joint university--industry appointment), only the primary (first-listed) affiliation was retained. This conservative approach provides a consistent annotation rule, though it may underestimate industry involvement and cross-sector collaboration for authors holding dual appointments. The country of each affiliation was determined based on the institutional headquarters location. A paper was classified as having industry involvement if at least one author's primary affiliation was a commercial or industrial entity (e.g., Google, Facebook/Meta, OpenAI, Microsoft, IBM, Tencent, ByteDance), as opposed to a university, government laboratory, or non-profit research institute.

\paragraph{Publication Venue.}
Each paper's publication venue was classified into one of five categories: Conference, Journal Article, Preprint, Book Chapter, or Workshop. No restrictions on venue type were imposed during data collection.

\paragraph{Limitations of the Annotation Process.}
All annotation was performed by a single researcher over a period of approximately six months, and no formal inter-rater reliability assessment was conducted. While this is a limitation, a single-rater design confers a compensating advantage: it ensures perfect internal consistency in the application of classification criteria across all 149 papers, avoiding the inter-rater disagreement noise that can arise when multiple annotators interpret ambiguous cases differently. Moreover, several of the annotation dimensions, particularly industry involvement (determined by whether an author's primary affiliation is a commercial entity) and embedding category (determined by the output type of the proposed method), are sufficiently objective that rater subjectivity is minimized. For more subjective classifications, such as the original-versus-variant distinction, ambiguous cases were resolved by prioritizing the paper's self-described contribution, and all decisions were verified against the original publications' abstracts, methodology sections, and stated contributions. Additionally, the restriction to English-language publications was not an explicit design choice but rather reflects the dominance of English in NLP research venues; our multi-source search strategy did not surface non-English papers meeting the inclusion criteria.

\section{Results}

\subsection{Descriptive Statistics}

\begin{figure}[h!]
    \centering
    \begin{tikzpicture}
        \begin{axis}[
            ybar,
            bar width=12pt,
            width=16cm,
            height=8cm,
            ymin=0,
            ymax=20,
            xlabel={Year},
            ylabel={Number of Papers},
            enlarge y limits={upper=0.2},
            symbolic x coords={1954,1972,1975,1990,1996,1999,2000,2003,2006,2007,2009,2010,2011,2012,2013,2014,2015,2016,2017,2018,2019,2020,2021,2022,2023,2024,2025},
            xtick=data,
            xticklabel style={rotate=45, anchor=east, font=\tiny},
            nodes near coords,
            nodes near coords style={font=\tiny},
            enlarge x limits=0.03,
            title={Distribution of Paper Publications per Year}
        ]
            \addplot[fill=light-purple] coordinates 
            {(1954,1) (1972,1) (1975,1) (1990,2) (1996,2) (1999,1) 
             (2000,1) (2003,2) (2006,1) (2007,2) (2009,1) (2010,1) 
             (2011,1) (2012,1) (2013,3) (2014,10) (2015,10) (2016,9) 
             (2017,14) (2018,16) (2019,15) (2020,19) (2021,4) (2022,6) 
             (2023,5) (2024,11) (2025,9)};
            \draw[red, dashed, line width=1.5pt] (axis cs:2020,0) -- (axis cs:2020,19);
            \node[above, red, font=\small\bfseries] at (axis cs:2020,20) {GPT-3 Introduced};
        \end{axis}
    \end{tikzpicture}
    \caption{Distribution of paper publications per year (1954-2025).}
    \label{fig:publications_per_year}
\end{figure}
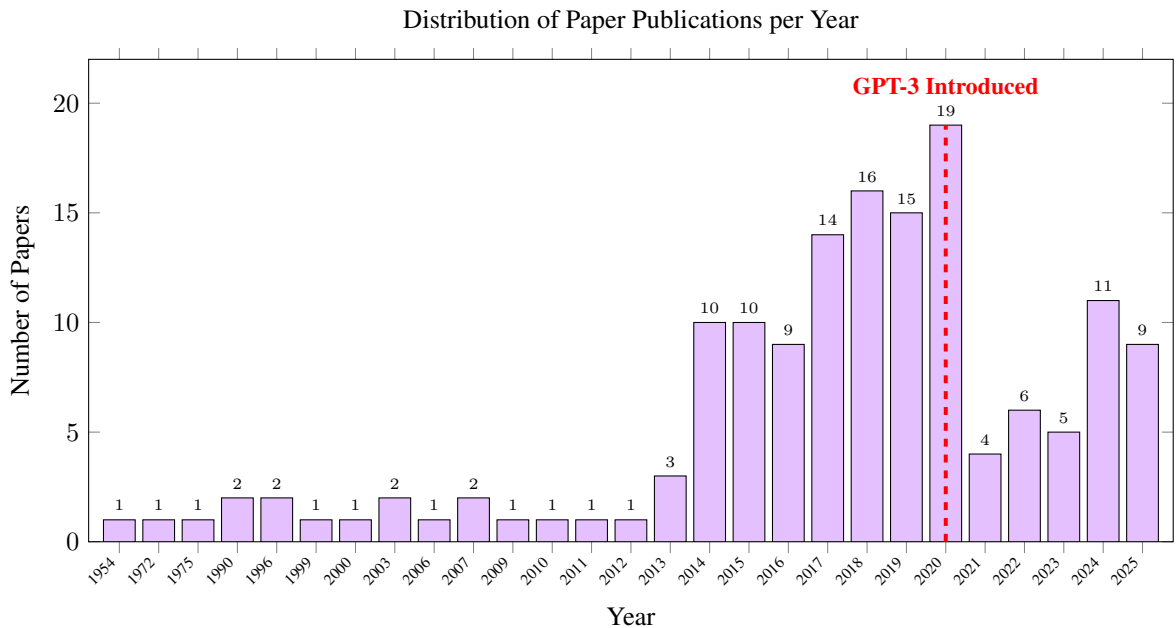

\noindent
Figure~\ref{fig:publications_per_year} presents the distribution of the number of word-embedding papers per year. We observe a clear peak in 2020, followed by a significant decline in subsequent years. The surge in 2020 can be explained by multiple factors. First, by this time, transformer-based architectures such as BERT \cite{devlin2019bert} and GPT-2 \cite{radford2019language} had rapidly gained traction in the NLP community, leading to a wave of studies exploring new embedding methods, adaptations, and enhancements. This excitement generated a spike in publications centered on embeddings, as researchers sought to build upon these breakthroughs.

\noindent
However, after 2020, the number of publications drops significantly. This decline is primarily attributable to a fundamental shift in the research landscape. After 2020, attention gradually moved from traditional word embeddings toward large language models (LLMs) and instruction-tuned systems such as GPT-3 \cite{brown2020language} and ChatGPT \cite{openai2022chatgpt}. As these paradigms became dominant, fewer papers focused specifically on word embeddings, reflecting the community's natural transition toward LLM-based approaches that rendered traditional embedding methods less central to cutting-edge research. Importantly, this decline does not indicate a diminished interest in semantic representation itself, but rather a shift in the locus of innovation. Embeddings have transitioned from being the primary \emph{product} of research, i.e., novel ways to construct vector representations, to a standardized \emph{component} within larger architectures, where the embedding layer typically functions as a fixed lookup table fed into a Transformer stack. In other words, the embedding problem has not been abandoned; it has been subsumed into the broader challenge of end-to-end language modeling.

\begin{figure}[h!]
    \centering
    \begin{tikzpicture}
        \begin{axis}[
            ybar,
            bar width=20pt,
            width=12cm,
            height=7cm,
            ymin=0,
            xlabel={Number of Authors},
            ylabel={Number of Papers},
            symbolic x coords={1,2,3,4,5,6,7,8,9,10+},
            xtick=data,
            nodes near coords,
            enlarge x limits=0.1,
            title={Distribution of Number of Authors}
        ]
            \addplot[fill=light-pink] coordinates 
            {(1,8) 
             (2,30) 
             (3,33)
             (4,24)
             (5,16)
             (6,17)
             (7,4)
             (8,7)
             (9,3)
             (10+,7)};
        \end{axis}
    \end{tikzpicture}
    \caption{Distribution of number of authors.}
    \label{fig:distribution_authors}
\end{figure}
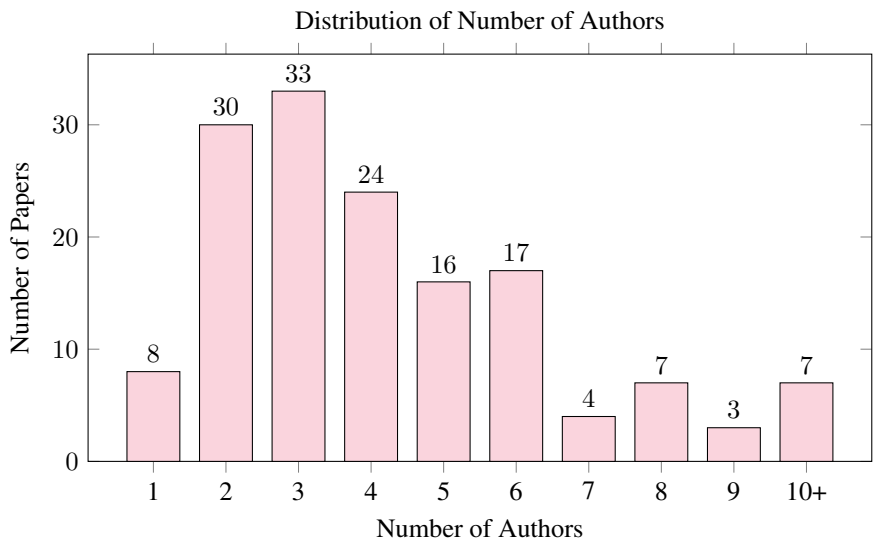

\noindent
Figure \ref{fig:distribution_authors} illustrates the distribution of the number of authors across the collected papers. The majority of studies are written by small to medium-sized teams, with a clear concentration around three to five authors. This pattern reflects the traditional research structure in NLP, where embedding-related projects were often conducted within a single lab or by small collaborations. As the number of authors increases beyond five, the frequency of publications gradually declines, indicating that larger collaborations are less common overall but still significant within the dataset.

\noindent
In our dataset, we found that papers with ten authors are disproportionately affiliated with major technology corporations or other industry research labs, rather than traditional university teams such as in these papers \cite{brown2020language,conneau2019unsupervised,liu2019roberta,cer2018universal,sun2019ernie,zaheer2020big}. This suggests that the larger author groups at this point are driven by industry-led or industry-dominated projects, where the scale of computational resources, infrastructure, and specialized roles requires broad internal collaboration.

\noindent
This finding corroborates earlier bibliometric claims in \cite{abdalla2023elephant, bosten2025conflicts} that industry–academic collaborations, and in some cases fully industry-led projects, are reshaping the authorship structure of NLP research. 

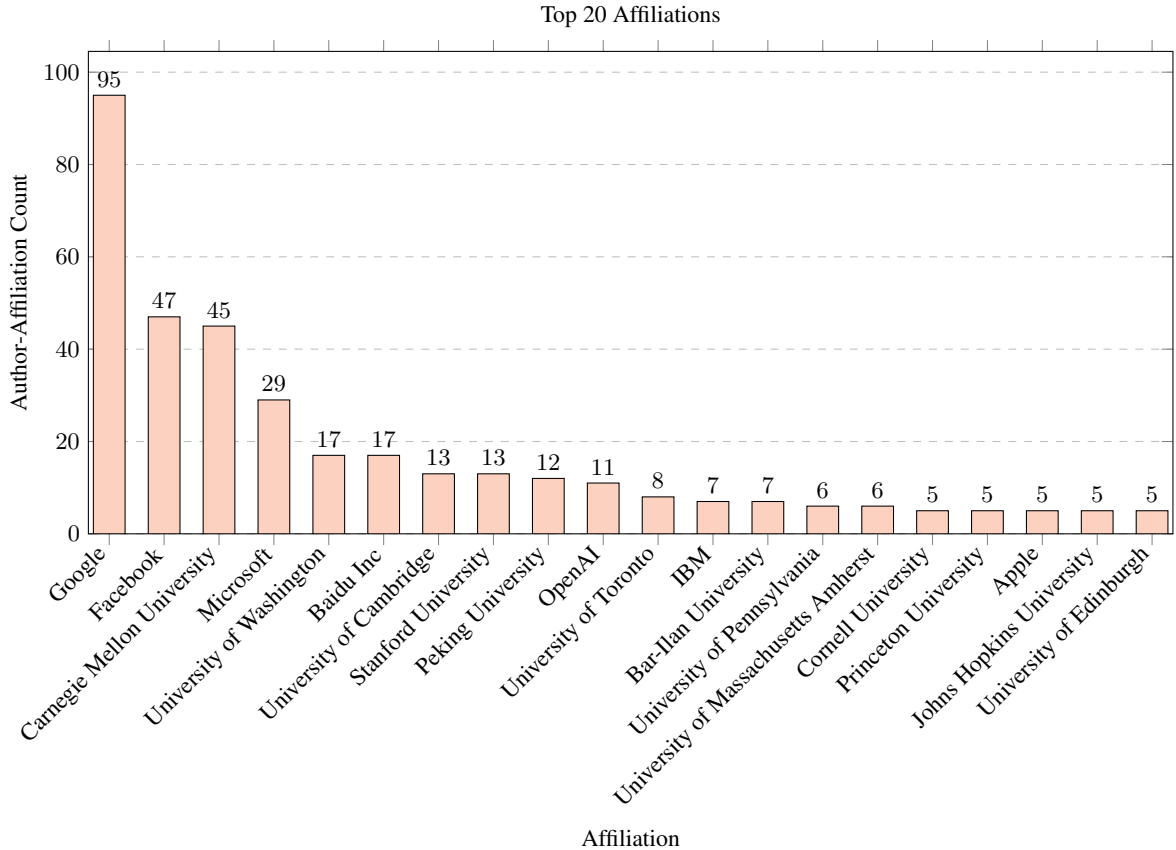
\begin{figure}[h!]
\centering
\footnotesize
\begin{tikzpicture}
\begin{axis}[
    ybar,
    bar width=12pt,
    width=16cm,
    height=8cm,
    ymin=0,
    xlabel={Affiliation},
    ylabel={Author-Affiliation Count},
    symbolic x coords={
        Google,Facebook,Carnegie Mellon University,Microsoft,University of Washington,
        Baidu Inc,University of Cambridge,Stanford University,Peking University,OpenAI,
        University of Toronto,IBM,Bar-Ilan University,University of Pennsylvania,
        University of Massachusetts Amherst,Cornell University,Princeton University,
        Apple,Johns Hopkins University,University of Edinburgh
    },
    xtick=data,
    xticklabel style={rotate=45, anchor=east, font=\footnotesize},
    nodes near coords,
    enlarge x limits=0.02,
    title={Top 20 Affiliations},
    ymajorgrids=true,
    grid style=dashed
]
\addplot[fill=light-red] coordinates
    {(Google,95)
     (Facebook,47)
     (Carnegie Mellon University,45)
     (Microsoft,29)
     (University of Washington,17)
     (Baidu Inc,17)
     (University of Cambridge,13)
     (Stanford University,13)
     (Peking University,12)
     (OpenAI,11)
     (University of Toronto,8)
     (IBM,7)
     (Bar-Ilan University,7)
     (University of Pennsylvania,6)
     (University of Massachusetts Amherst,6)
     (Cornell University,5)
     (Princeton University,5)
     (Apple,5)
     (Johns Hopkins University,5)
     (University of Edinburgh,5)};
\end{axis}
\end{tikzpicture}
\caption{Top 20 affiliations ranked by author-affiliation count.}
\label{fig:distribution_affiliations}
\end{figure}

\noindent
Figure \ref{fig:distribution_affiliations} presents the distribution of the top 20 affiliations represented across the 149 papers in our dataset. Google leads overwhelmingly with 95 affiliations, followed by Facebook (47), Carnegie Mellon University (45), and Microsoft (29). Other major contributors include the University of Washington (17), Baidu Inc (17), the University of Cambridge (13), Stanford University (13), Peking University (12), and OpenAI (11). Several additional universities and industry research groups, such as the University of Toronto, IBM, and Apple, appear among the remaining affiliations, reflecting both academic and corporate engagement in advancing word-embedding research.

\begin{figure}[h!]
    \centering
    \begin{tikzpicture}
        \begin{axis}[
            ybar,
            bar width=10pt,
            width=15cm,
            height=8cm,
            ymin=0,
            xlabel={Nation},
            ylabel={Number of Papers},
            symbolic x coords={ USA,China,UK,Germany,Canada,Israel,Japan,France,Switzerland,Singapore,Taiwan,India,Korea,Denmark,Portugal,Netherlands,Sweden,Russia,Poland,Czechia
            },
            xtick=data,
            xticklabel style={rotate=45, anchor=east, font=\footnotesize},
            nodes near coords,
            enlarge x limits=0.02,
            title={Top 20 Nations},
            ymajorgrids=true,
            grid style=dashed
        ]
            \addplot[fill=light-yellow] coordinates
            {(USA,378)
             (China,86)
             (UK,25)
             (Germany,19)
             (Canada,14)
             (Israel,11)
             (Japan,11)
             (France,9)
             (Switzerland,7)
             (Singapore,6)
             (Taiwan,5)
             (India,4)
             (Korea,4)
             (Denmark,4)
             (Portugal,4)
             (Netherlands,3)
             (Sweden,3)
             (Russia,3)
             (Poland,3)
             (Czechia,3)};
        \end{axis}
    \end{tikzpicture}
    \caption{Top 20 nations ranked by author-affiliation count.}
    \label{fig:distribution_nations}
\end{figure}
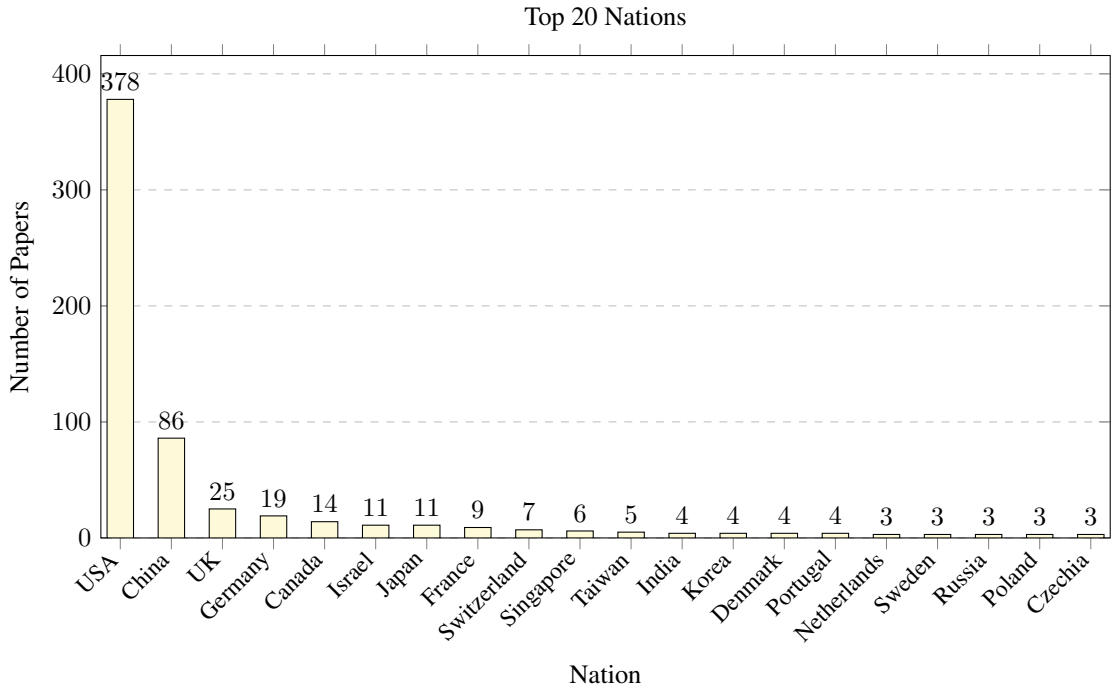

\noindent
Figure \ref{fig:distribution_nations} illustrates the distribution of papers by country affiliation, highlighting the top 20 nations contributing to word-embedding research. The United States dominates with 378 affiliations, followed by China (86), the UK (25), Germany (19), and Canada (14). Mid-tier contributors include Israel (11), Japan (11), France (9), Switzerland (7), and Singapore (6). Other active countries such as Taiwan, India, Korea, Denmark, Portugal, the Netherlands, Sweden, Russia, Poland, and Czechia also make meaningful contributions. (Note: Scottish affiliations are counted under the UK.)

\noindent
Taken together, Figures~\ref{fig:distribution_affiliations} and~\ref{fig:distribution_nations} reveal a pronounced concentration of word embedding research in a small number of resource-rich environments. The United States dominates both institutional and national rankings, driven by a strong ecosystem of major technology companies (Google, Facebook, Microsoft), leading universities (Carnegie Mellon, Stanford, University of Washington), and substantial government and private-sector funding for AI research. China's growing presence correlates with state-supported initiatives in machine learning, particularly through firms such as Baidu and Tencent and universities such as Peking University. The lower representation from other nations suggests potential barriers, including limited computational infrastructure, smaller research communities, and language biases in datasets, that could restrict diverse perspectives in embedding techniques and limit progress on multilingual representations.

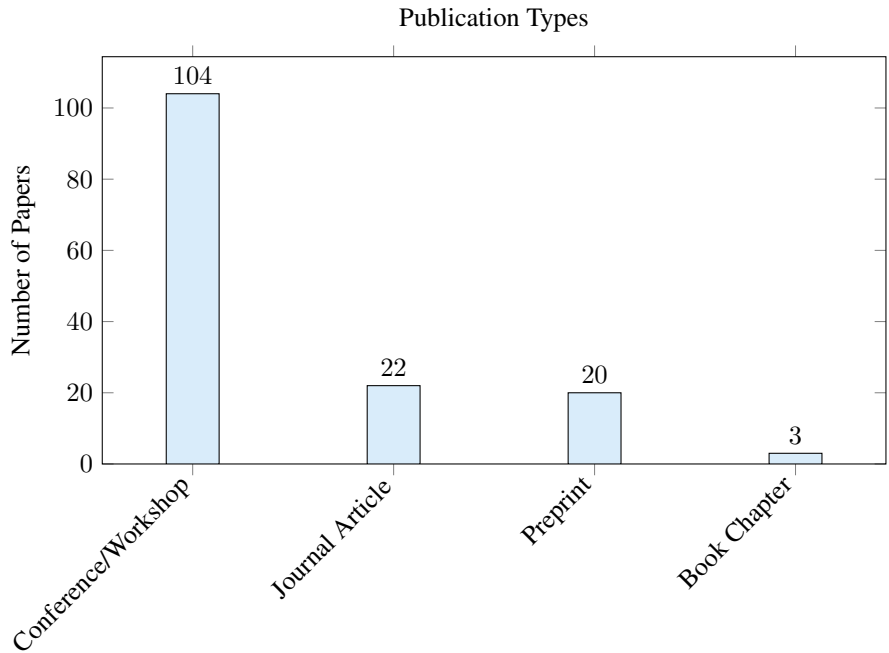
\begin{figure}[h!]
    \centering
    \begin{tikzpicture}
        \begin{axis}[
            ybar,
            bar width=20pt,
            width=12cm,
            height=7cm,
            ymin=0,
            ylabel={Number of Papers},
            symbolic x coords={Conference/Workshop,Journal Article,Preprint,Book Chapter,Workshop},
            xtick=data,
            xticklabel style={rotate=45, anchor=east},
            nodes near coords,
            enlarge x limits=0.15,
            title={Publication Types}
        ]
            \addplot[fill=light-blue] coordinates 
            {(Conference/Workshop,104) 
             (Journal Article,22)
             (Preprint,20) 
             (Book Chapter,3)};
        \end{axis}
    \end{tikzpicture}
    \caption{Number of papers by publication type.}
    \label{fig:publication_types}
\end{figure}

\noindent
Figure~\ref{fig:publication_types} illustrates the distribution of publication types across the collected papers. The majority of word embedding studies, 103 out of 149, are published through conferences, plus one additional workshop paper, yielding a combined total of 104 conference/workshop publications. This dominance underscores the iterative, collaborative environment of conferences, which prioritize timely exchange of cutting-edge ideas in computational linguistics. 

\noindent
Journal articles represent the second-largest category, with 22 entries. These typically offer more comprehensive or longitudinal treatments of embedding methods compared to conference publications. Preprints account for the next 20 papers, often disseminated via arXiv, where they achieve rapid visibility. Notably, preprints in this field are not necessarily preliminary works; many introduce groundbreaking ideas that shape the field prior to or alongside peer-reviewed publication, as seen in contributions such as \cite{wang2022text,beltagy2020longformer}. Meanwhile, book chapters (3) remain relatively rare within the dataset.

\begin{figure}[htbp]
    \centering
    \begin{tikzpicture}
        \begin{axis}[
            ybar,
            bar width=20pt,
            width=12cm,
            height=7cm,
            ymin=0,
            ylabel={Number of Papers},
            symbolic x coords={Static Word Embeddings,Contextual Word Embeddings,Sentence/Document Embedding,Statistical Representations},
            xtick=data,
            xticklabel style={rotate=45, anchor=east},
            nodes near coords,
            enlarge x limits=0.2,
            title={Frequency of Embedding Method Categories}
        ]
            \addplot[fill=light-green] coordinates 
            {(Static Word Embeddings,50) 
             (Contextual Word Embeddings,45) 
             (Sentence/Document Embedding,34) 
             (Statistical Representations,20)};
        \end{axis}
    \end{tikzpicture}
    \caption{The frequency of embedding method categories.}
    \label{fig:frequency_embedding_method}
\end{figure}
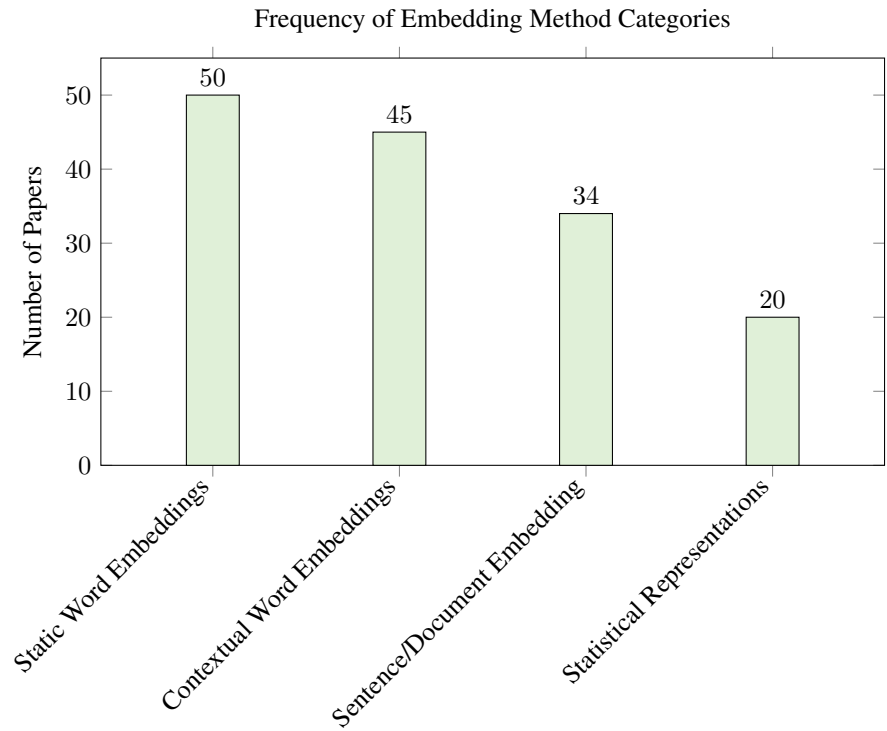

\noindent
As shown in Figure~\ref{fig:frequency_embedding_method}, static word embeddings represent the largest category, with 50 papers, closely followed by contextual embeddings at 45 papers. Sentence and document-level embeddings account for 34 studies, while statistical representations are the least common with only 20. This distribution highlights the community’s strong emphasis on neural-based approaches, especially static and contextual models, over earlier statistical methods.

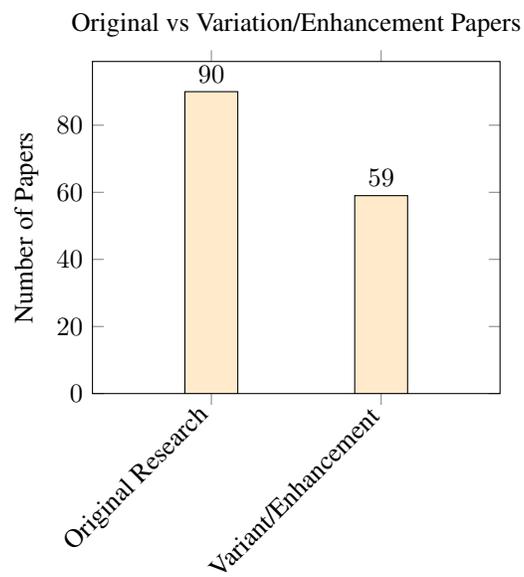
\begin{figure}[htbp]
    \centering
    \begin{tikzpicture}
        \begin{axis}[
            ybar,
            bar width=20pt,
            width=7cm,
            height=6cm,
            ymin=0,
            ylabel={Number of Papers},
            symbolic x coords={Original Research,Variant/Enhancement},
            xtick=data,
            xticklabel style={rotate=45, anchor=east},
            nodes near coords,
            enlarge x limits=0.7,
            title={Original vs Variation/Enhancement Papers}
        ]
            \addplot[fill=light-orange] coordinates 
            {(Original Research,90) 
             (Variant/Enhancement,59)};
        \end{axis}
    \end{tikzpicture}
    \caption{Original vs.\ Variation/Enhancement papers.}
    \label{fig:original_vs_variation}
\end{figure}

\noindent
Figure~\ref{fig:original_vs_variation} compares original research papers to those presenting variants or enhancements. The results show that original contributions dominate the field, with 90 papers, whereas 59 studies focus on extending or refining existing techniques. This suggests that while innovation through entirely new methods has been central, a considerable body of work is also dedicated to iterative improvements on established approaches.

\begin{figure}[h!]
    \centering
    \includegraphics[width=\textwidth]{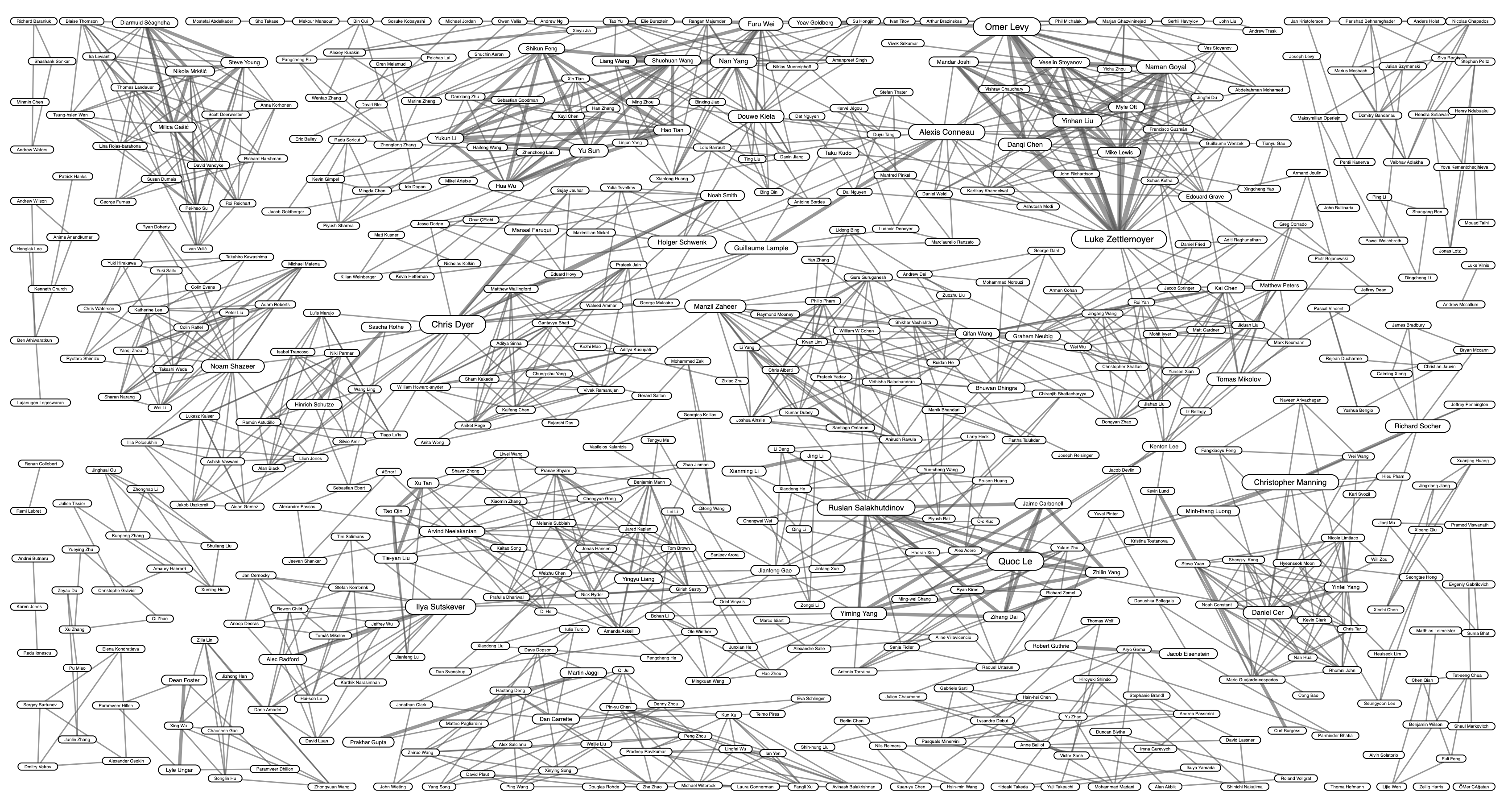}
    \caption{Coauthor network for the word-embedding literature. Node size encodes the number of publications per author, edge thickness encodes coauthorship frequency, and shaded regions indicate densely connected author clusters.}
    \label{fig:coauthor_network}
\end{figure}

\begin{figure}[h!]
    \centering
    \includegraphics[width=\textwidth]{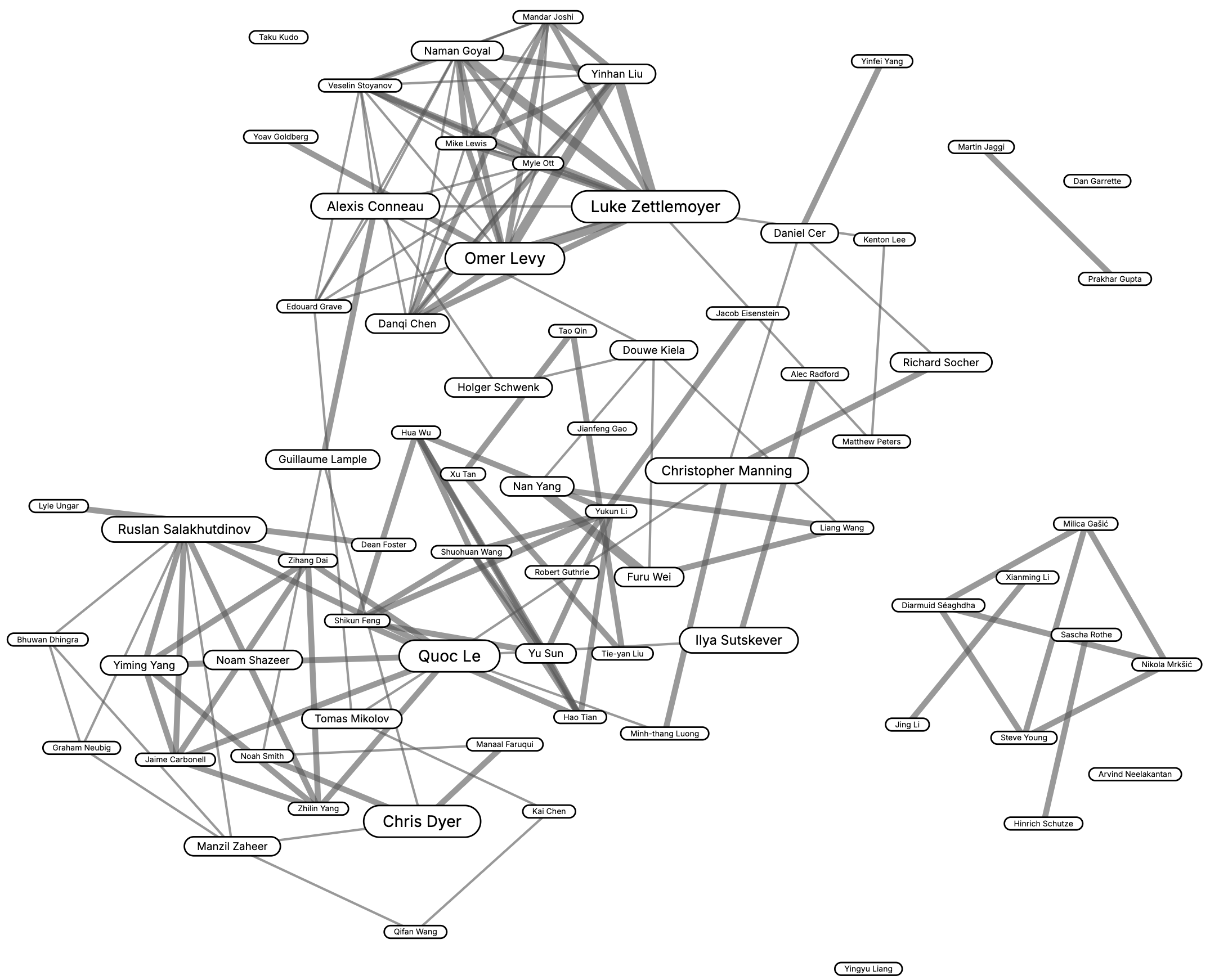}
    \caption{Coauthor graph with authors having at least 2 publications in the dataset. This filtered view highlights the core research communities.}
    \label{fig:coauthor_network_filtered}
\end{figure}

\noindent
Figure~\ref{fig:coauthor_network} presents the coauthor network constructed from the 149 papers in our dataset. Each node corresponds to an individual author, with node size proportional to the number of publications in the corpus. Edges represent coauthorship, and thicker edges indicate authors who have collaborated on multiple word-embedding papers. Cluster regions group densely connected authors, revealing collaboration communities that often align with specific institutions, labs, or research programs. To further clarify these core communities, Figure~\ref{fig:coauthor_network_filtered} filters the network to display only authors with at least two publications, highlighting the sustained research groups that drive long-term methodological development in the field. It is important to note that due to the high density of the network, node sizes and edge widths in both figures are scaled for visualization clarity rather than representing precise counts.

\noindent
To understand how collaboration dynamics evolved over time, we partitioned the dataset into three periods based on major paradigm shifts in the field: early foundations (1954--2012, $n = 18$), the neural embedding era (2013--2019, $n = 77$), and the LLM transition period (2020--2025, $n = 54$). The first period spans from Harris's foundational work on distributional structure \cite{harris1954distributional} through the final year before Word2Vec, capturing the era of sparse statistical methods such as Bag-of-Words, TF-IDF, and LSA. The second period begins with the release of Word2Vec \cite{mikolov2013efficient} in 2013, which fundamentally transformed the field by introducing efficient dense neural embeddings, and extends through 2019, encompassing the rise of static embeddings (GloVe, FastText) and early contextual models (ELMo, BERT). The third period starts in 2020, marked by the release of GPT-3 \cite{brown2020language}, which signaled the shift toward massive language models where embeddings became internal representations within end-to-end systems rather than standalone components.

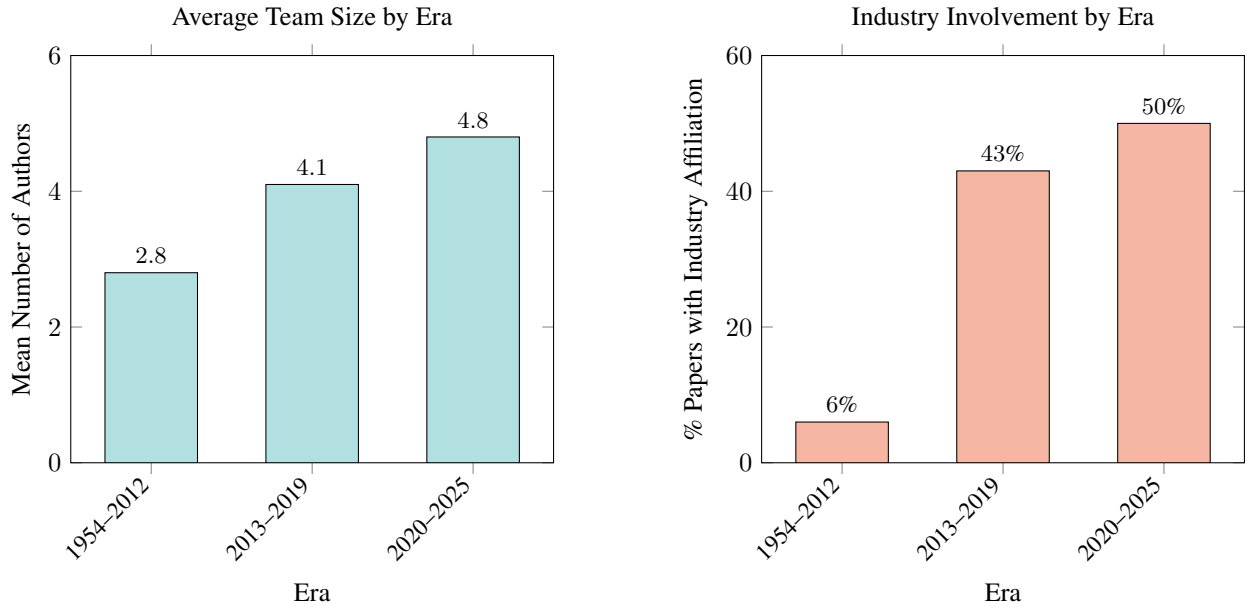
\begin{figure}[h!]
    \centering
    \begin{tikzpicture}
        \begin{axis}[
            ybar,
            bar width=35pt,
            width=8cm,
            height=7cm,
            ymin=0,
            ymax=6,
            xlabel={Era},
            ylabel={Mean Number of Authors},
            symbolic x coords={1954--2012,2013--2019,2020--2025},
            xtick=data,
            xticklabel style={rotate=45, anchor=east, font=\footnotesize},
            nodes near coords,
            nodes near coords style={font=\small},
            enlarge x limits=0.25,
            title={Average Team Size by Era}
        ]
            \addplot[fill=light-teal] coordinates
            {(1954--2012,2.8) (2013--2019,4.1) (2020--2025,4.8)};
        \end{axis}
    \end{tikzpicture}
    \hfill
    \begin{tikzpicture}
        \begin{axis}[
            ybar,
            bar width=35pt,
            width=8cm,
            height=7cm,
            ymin=0,
            ymax=60,
            xlabel={Era},
            ylabel={\% Papers with Industry Affiliation},
            symbolic x coords={1954--2012,2013--2019,2020--2025},
            xtick=data,
            xticklabel style={rotate=45, anchor=east, font=\footnotesize},
            nodes near coords={\pgfmathprintnumber\pgfplotspointmeta\%},
            nodes near coords style={font=\small},
            enlarge x limits=0.25,
            title={Industry Involvement by Era}
        ]
            \addplot[fill=light-coral] coordinates
            {(1954--2012,6) (2013--2019,43) (2020--2025,50)};
        \end{axis}
    \end{tikzpicture}
    \caption{Evolution of collaboration dynamics across three eras. Left: mean number of authors per paper. Right: percentage of papers with at least one industry affiliation.}
    \label{fig:collaboration_dynamics}
\end{figure}

\noindent
As shown in Figure~\ref{fig:collaboration_dynamics}, these three eras exhibit distinct collaboration patterns. The average number of authors per paper increased from 2.8 in the early period to 4.1 during the neural era and 4.8 in the LLM period. Industry involvement also rose dramatically: only 6\% of early papers included industry affiliations, compared to 43\% during the neural era and 50\% in the LLM period. These descriptive patterns are further corroborated by formal statistical testing in the Before vs. After GPT-3: Era Comparison section, confirming that word embedding research has shifted from small academic teams toward larger, industry-involved collaborations requiring substantial computational resources.

\subsection{Analysis on Specific Methods}

Although our dataset comprises 149 papers spanning seven decades, it is impractical to provide a detailed analysis of every individual contribution. Therefore, for the qualitative analysis in this section, we have selected specific techniques that serve as \textbf{archetypes} for their respective categories. Specifically, these selected papers satisfy three criteria: (1) they are \textbf{seminal works} that fundamentally defined a new approach (e.g., Word2Vec for prediction-based embeddings); (2) they represent a \textbf{paradigm shift} or a "first" in the field (e.g., FastText for subword information); or (3) they are \textbf{highly influential foundational works} that serve as parents to numerous derivative studies. By focusing on these representative methods, we aim to trace the structural evolution of the field rather than enumerating every minor variation.

\begin{figure}[ht]
    \centering
    \includegraphics[width=1.0\linewidth]{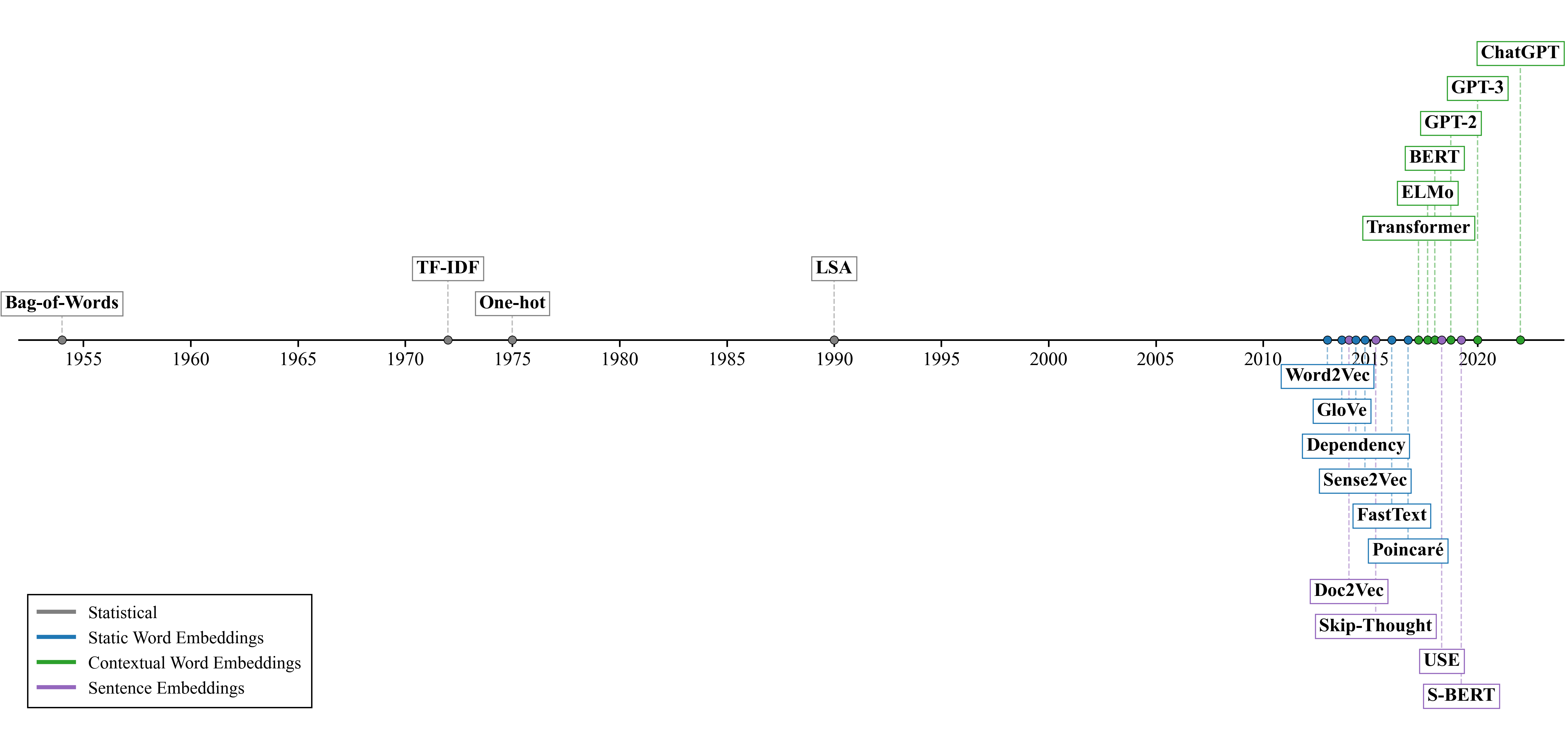}
    \caption{Timeline of major word embedding techniques (1954–2022). The field evolved in four distinct phases: (1) \textbf{Statistical} methods (grey) focused on frequency counts; (2) \textbf{Static} embeddings (blue) introduced dense vector learning; (3) \textbf{Contextual} models (green) achieved dynamic representations culminating in LLMs; and (4) \textbf{Sentence} embeddings (purple) extended these principles to longer text units.}
    \label{fig:timeline}
\end{figure}

As illustrated in Figure \ref{fig:timeline}, the evolution of these techniques follows a clear trajectory from sparse statistical representations to dense, context-aware neural models. The detailed analysis of these four categories follows below.

\subsubsection{Statistical Representation-Based Methods}

\textbf{One-hot Encoding}. Technically, one-hot vectors, which can be traced to early vector space models \cite{salton1975vector}, are not a learned "word embedding" but a sparse indicator representation: each vocabulary item is a basis vector with a single 1 and zeros elsewhere, so the geometry encodes identity only and every pair of distinct words is equally distant. As a representation it is extremely sparse and very high-dimensional, provides no notion of semantic similarity, and cannot generalize to unseen tokens; its main modern role is practical, serving as indices into learned embedding tables or as target labels in classifiers—rather than as standalone features. Precisely because it is so limited, one-hot usefully highlights what subsequent methods add: parameter sharing, distributional similarity, and the need for weighting or dimensionality reduction to manage sparsity.

\textbf{Bag-of-Words}. Bag-of-Words (BoW), drawing on principles from Harris's distributional hypothesis \cite{harris1954distributional}, aggregates one-hot indicators over a span (typically a document) into a term-frequency vector, optionally extending the vocabulary with n-grams and normalizing by length. Despite discarding word order and syntax, BoW remains a strong, interpretable baseline for retrieval and text classification because it captures salient lexical cues and makes feature importance transparent. Its weaknesses are equally clear: very sparse, high-dimensional vectors dominated by frequent words, brittleness to morphology and out-of-vocabulary items, and no modeling of compositional structure. In practice, BoW is commonly paired with vocabulary pruning, reweighting (e.g., TF-IDF \cite{sparck1972statistical}), and dimensionality reduction (e.g., truncated SVD/LSA \cite{deerwester1990indexing,landauer1997solution,halko2011finding}) to emphasize informative terms and control sparsity. Despite its simplicity, it is widely recognized that the core BoW principle---representing text as an unordered collection of token indicators---persists in modern architectures: Transformer models still begin by mapping each input token to a vector via a lookup table, and pooling strategies that aggregate token embeddings into a single sequence representation echo the same ``bag'' abstraction at a higher level of representation.

\textbf{TF-IDF}. Term Frequency--Inverse Document Frequency (TF-IDF) \cite{sparck1972statistical} improves on the traditional Bag-of-Words model by preserving the same sparse vector representation while assigning more informative weights to each term. Instead of treating every word count equally, TF-IDF increases the importance of words that are characteristic of a specific document and decreases the weight of extremely common words that appear across many documents.

A standard TF-IDF weight for a word $w$ in a document $d$ is defined as:
\[
\text{TF-IDF}(w,d) = \text{TF}(w,d) \cdot \log\!\left(\frac{N}{\text{DF}(w)+1}\right),
\]
where $\text{TF}(w,d)$ is the term frequency of $w$ in $d$, $\text{DF}(w)$ is the number of documents containing $w$, and $N$ is the total number of documents.

Compared to Bag-of-Words, TF--IDF introduces the notion of informativeness: rare but meaningful words receive larger weights, while ubiquitous words are downweighted toward zero. This weighting makes TF--IDF a stronger baseline for information retrieval and early text classification. However, it still ignores word order and does not provide distributed or continuous semantic representations, consequently its ability to capture deeper linguistic structure remains limited compared to modern embedding-based methods. As commonly noted in the literature, the core principle of TF-IDF, assigning variable weights to tokens based on their informativeness, can be seen as a conceptual precursor to the attention mechanism in Transformers \cite{vaswani2017attention}, which dynamically computes relevance-based weights over input tokens. Where TF-IDF uses corpus-level statistics to determine importance, self-attention learns context-dependent importance through query-key interactions, but the underlying intuition that ``not all tokens matter equally'' is shared across both paradigms.

\textbf{Latent Semantic Analysis (LSA)}. While TF-IDF addresses term weighting, it does not solve the problem of synonymy (different words having similar meanings) or polysemy. Latent Semantic Analysis (LSA), introduced by Deerwester et al.\ (1990) \cite{deerwester1990indexing}, addresses this by applying Singular Value Decomposition (SVD) to the term-document matrix. By truncating the decomposition to keep only the top $k$ singular values, LSA projects the high-dimensional sparse data into a lower-dimensional "latent semantic space." In this reduced space, terms that co-occur in similar contexts appear near each other, capturing semantic similarities that raw term overlap misses. LSA represents the first major successful attempt at distributional semantic modeling via dimensionality reduction, establishing the core idea that a low-rank approximation of occurrence data can uncover hidden conceptual relationships. As established in the literature---most notably by Levy and Goldberg \cite{levy2014neural}, who showed that Word2Vec's Skip-Gram implicitly factorizes a PMI matrix---LSA is the direct mathematical precursor to the dense vector spaces of Word2Vec and GloVe: all three methods seek a low-dimensional space that preserves distributional similarity, but where LSA achieves this through linear algebraic decomposition, neural embedding methods replace the decomposition with stochastic gradient optimization over a prediction or reconstruction objective.

\textbf{Neural Probabilistic Language Model}. Bengio et al.\ (2003) \cite{bengio2003neural} introduced a pivotal bridge between classical statistical methods and modern neural embeddings. Their neural probabilistic language model learned distributed word representations as part of a neural network trained to predict the next word in a sequence. Each word was mapped to a continuous vector, and these vectors were jointly learned with the model's parameters. This approach demonstrated that dense, low-dimensional word representations could be learned end-to-end through backpropagation, directly optimizing a language modeling objective. The work laid the conceptual foundation for Word2Vec and subsequent neural embedding methods, showing that learned embeddings could capture rich semantic and syntactic regularities while enabling generalization to unseen word combinations.

\subsubsection{Static Word Embeddings}

\textbf{Word2Vec}. Word2Vec, introduced by Mikolov et al.\ (2013) \cite{mikolov2013efficient}, marked a major shift from sparse statistical representations toward dense, distributed word vectors learned from raw text. It consists of two predictive architectures, Continuous Bag-of-Words (CBOW) and Skip-Gram, designed to learn embeddings by maximizing the probability of observing context words around a target. Through efficient training strategies such as hierarchical softmax and negative sampling, Word2Vec scales to extremely large corpora and produces vector spaces where semantic and syntactic regularities emerge naturally, enabling analogical reasoning (e.g., ``$\textit{king} - \textit{man} + \textit{woman} \approx \textit{queen}$''). Its strengths include computational efficiency, strong representational quality, and broad applicability across NLP tasks. However, Word2Vec assigns one vector per word type, making it unable to model polysemy, domain-specific word senses, or context-dependent meaning. Moreover, it relies on small local windows and does not incorporate global corpus statistics, leading to instability in rare-word representations and sensitivity to hyperparameters. These limitations laid the foundation for improved approaches, most notably GloVe, which explicitly incorporates global co-occurrence statistics to stabilize and refine embedding quality.

\textbf{GloVe}. GloVe (Global Vectors), proposed by Pennington, Socher, and Manning (2014) \cite{pennington2014glove}, builds on Word2Vec by integrating global corpus statistics into the embedding learning process. Instead of predicting local contexts, GloVe factorizes a word--context co-occurrence matrix using a weighted least squares objective derived from ratios of co-occurrence probabilities. This design captures both local semantic patterns and global distributional structure, producing more stable embeddings and smoother vector spaces compared to purely predictive models. GloVe excels in analogy tasks and provides more interpretable linear substructures due to its grounding in log co-occurrence ratios. It also mitigates some weaknesses of Word2Vec, particularly sensitivity to local windows and instability in rare-word contexts, through its principled use of whole-corpus statistics. Nevertheless, GloVe remains, like Word2Vec, a static embedding model, mapping each word type to a single vector regardless of context. It also requires building and storing large co-occurrence matrices, which can be computationally costly for very large corpora. These remaining challenges motivated the development of FastText, which further enriches static embeddings by incorporating subword information to better model morphology and rare or unseen words.

\textbf{FastText}. FastText, introduced by Bojanowski et al.\ (2017) \cite{bojanowski2017enriching}, extends static word embeddings by incorporating subword-level information, thereby addressing key limitations in both Word2Vec and GloVe. In FastText, each word is represented as a bag of character $n$-grams, and its embedding is computed as the sum of its subword vectors. This design enables the model to capture morphological patterns (for example, prefixes and suffixes), systematically represent related word forms (for example, \emph{walk}, \emph{walking}, \emph{walker}), and generate meaningful vectors for rare or even out-of-vocabulary words, capabilities that neither Word2Vec nor GloVe possess. FastText retains the training efficiency of the Skip-Gram architecture while significantly improving performance in morphologically rich languages and low-resource settings. However, it still inherits the major limitation of static embeddings: a word's meaning does not change with context, preventing it from distinguishing between different senses of polysemous terms (for example, \emph{bank} as a financial institution versus \emph{river bank}). Despite this, FastText represents the most refined form of static word embedding before the emergence of contextual models such as ELMo and BERT, offering an effective bridge between predictive neural approaches and the richer linguistic modeling that followed.

\textbf{Dependency-Based Word Embeddings}. A fundamental limitation of standard window-based models is their reliance on linear proximity, which often conflates topical relatedness with functional similarity. For instance, "Hogwarts" (a school) appears frequently near "Dumbledore" (its headmaster), creating a strong topical association. However, functionally, "Hogwarts" is a school, similar to "Sunnydale" (another fictional school), not a person. Levy and Goldberg (2014) \cite{levy2014dependency} addressed this by defining context through syntactic dependencies rather than linear windows. By using dependency parses to identify context words, this method filters out irrelevant neighbors and captures non-local relationships. The resulting embeddings exhibit a qualitative shift: whereas linear bag-of-words contexts cluster words by broad topical association (e.g., grouping "Hogwarts" with characters like "Dumbledore"), dependency-based contexts cluster words by functional role (e.g., grouping "Hogwarts" with other institutions like "Sunnydale" or "Yale"). This distinction highlights how the definition of "context" fundamentally shapes the semantic properties of the learned vector space.

\textbf{Poincaré Embeddings}. A critical issue with standard embeddings is their use of Euclidean space, which is inherently flat and cannot efficiently embed hierarchical structures (like taxonomies) without significant distortion. Nickel and Kiela (2017) \cite{nickel2017poincare} argued that this geometry is suboptimal for representing hypernymy relations (e.g., \emph{animal} $\rightarrow$ \emph{mammal} $\rightarrow$ \emph{dog}). They introduced Poincaré embeddings, which learn representations in a hyperbolic space. In this geometry, the volume of space increases exponentially with distance from the origin, allowing for a natural representation of hierarchies where root concepts are placed near the origin and leaf nodes are pushed toward the boundary. This approach allows for high-quality representations of complex hierarchies with significantly fewer dimensions than required in Euclidean space.

\textbf{Sense2Vec}. The most significant bottleneck for static embeddings is polysemy: they collapse all meanings of a word (e.g., "duck" as a bird vs. "duck" as a verb) into a single vector, losing distinct semantic senses. To address this without transitioning to full contextual models, Trask et al. (2015) \cite{trask2015sense2vec} introduced Sense2Vec, a supervised modification of Word2Vec that incorporates Part-of-Speech (POS) tags directly into the tokenization process. Instead of learning a single vector for "duck", Sense2Vec learns distinct vectors for "duck|NOUN" and "duck|VERB". By treating these as separate tokens, the model can disentangle multiple senses of a word provided they are syntactically distinct, representing a significant practical enhancement for static embedding pipelines.

Despite these innovations, the fundamental constraint of a fixed vector per word proved insurmountable for complex language understanding. Consequently, the field eventually moved away from static models entirely in favor of dynamic, contextual architectures like ELMo and BERT \cite{camacho2018from}.

\subsubsection{Contextual Word Embeddings}

\textbf{Transformer}. The transition to contextual embeddings was fundamentally enabled by the Transformer architecture, introduced by Vaswani et al.\ (2017) \cite{vaswani2017attention}. Although originally proposed for machine translation, the Transformer introduced the "Self-Attention" mechanism, which computes representations for each word by attending to all other words in the sequence simultaneously, regardless of distance. This broke the sequential bottleneck of Recurrent Neural Networks (RNNs) and LSTMs, allowing for massive parallelization and the modeling of long-range dependencies. While the original paper focused on encoder-decoder translation, its encoder block became the basis for BERT, and its decoder block became the basis for GPT. The Transformer is the architectural foundation of the entire modern era of NLP, shifting the paradigm from processing words sequentially to processing entire contexts holistically.

\textbf{ELMo}. ELMo, introduced by Peters et al.\ (2018) \cite{peters2018deep}, represents the first major step toward contextual word embeddings by producing a different vector for the same word depending on its surrounding context. Technically, ELMo is built on a deep bidirectional LSTM language model \cite{hochreiter1997long} trained at the character level, allowing it to capture morphology and long-range dependencies more effectively than earlier static embedding methods. Each ELMo embedding is a learned weighted combination of multiple internal LSTM layers, enabling lower layers to encode syntactic patterns while higher layers capture semantic nuances. This architecture fills a key gap left by static embeddings such as Word2Vec, GloVe, and FastText: the inability to model polysemy or context-dependent meaning. Although ELMo improves performance across many NLP tasks, its recurrent structure is computationally slower and less parallelizable than self-attention-based models. These limitations motivated the transition toward Transformer encoders \cite{vaswani2017attention}, as popularized by BERT, which offer more scalable contextualization and richer bidirectional representations.

\textbf{BERT}. BERT (Bidirectional Encoder Representations from Transformers), proposed by Devlin et al.\ (2019) \cite{devlin2019bert}, extends contextual word embeddings by replacing recurrent architectures with a multi-layer Transformer encoder \cite{vaswani2017attention}. BERT is trained using two self-supervised objectives: Masked Language Modeling (MLM) and Next Sentence Prediction (NSP), which enable it to learn bidirectional dependencies and model context holistically. Self-attention allows BERT to compute token relationships in parallel, giving it substantial advantages over ELMo’s sequential LSTM backbone and producing richer global context representations. BERT's strengths include strong performance on a wide variety of discriminative NLP tasks, efficient fine-tuning, and robust contextual embeddings. However, BERT is not inherently generative and relies on fixed-length input windows, and its masked objective does not define a full probabilistic model of text. These gaps opened the door for GPT, which uses a Transformer decoder architecture to model text autoregressively and support more flexible generation.

\textbf{GPT}. GPT-style models, beginning with Radford et al.\ (2018) \cite{radford2018improving} and extended in GPT-2 (2019) \cite{radford2019language}, shift contextual embeddings into a fully generative setting by training a unidirectional decoder-only Transformer  \cite{vaswani2017attention} to predict each token from its left context using causal self-attention. This autoregressive objective directly models the probability distribution of text sequences, enabling GPT to excel at long-form text generation, story continuation, and in-context learning. Technically, GPT learns hierarchical representations across decoder layers, where lower layers encode local syntactic features while higher layers capture semantics, discourse structure, and reasoning patterns. GPT fills critical gaps left by BERT: it naturally supports coherent generation without masking strategies, uses prompting instead of task-specific output heads, and exhibits strong few-shot and zero-shot behavior. Nonetheless, GPT lacks explicit bidirectional context, requires significant computational resources to train, and is sensitive to prompt design. Despite these limitations, GPT demonstrates how autoregressive Transformer embeddings can generalize far beyond word meaning to support broad linguistic and reasoning capabilities.

\subsubsection{Sentence and Document Embeddings}

\textbf{Doc2Vec}. While Word2Vec revolutionized word representations, it lacked a native mechanism for representing variable-length texts like sentences or documents. Le and Mikolov (2014) \cite{le2014distributed} addressed this with Doc2Vec (or Paragraph Vector), which extends the Word2Vec architecture by adding a unique "paragraph vector" that acts as a memory of the topic of the document. In the PV-DM (Distributed Memory) model, this paragraph vector is concatenated with word vectors to predict the next word, effectively conditioning the prediction on the document's global context. Doc2Vec demonstrated that simple vector arithmetic could generalize to documents (e.g., matching a review to its sentiment) and established a baseline for unsupervised document representation. However, it requires a new inference step (gradient descent) to compute a vector for an unseen document, making it slower at test time compared to purely compositional methods.

\textbf{Skip-Thought Vectors}. Moving beyond bag-of-words approaches, Kiros et al. (2015) \cite{kiros2015skip} proposed Skip-Thought Vectors, which generalized the skip-gram objective to the sentence level. Using an encoder-decoder architecture, Skip-Thought encodes a central sentence and tries to predict the sentences immediately preceding and following it. This objective forces the model to capture semantic and syntactic properties required to reconstruct the surrounding discourse. While Skip-Thought produced highly generic sentence representations useful for diverse downstream tasks, it was computationally expensive to train and extremely slow for inference due to its heavy reliance on recurrent neural networks (RNNs).

\textbf{Universal Sentence Encoder}. To provide a more efficient and versatile standard for sentence embeddings, Cer et al. (2018) \cite{cer2018universal} introduced the Universal Sentence Encoder (USE). USE offers two distinct architectures to trade off between accuracy and computational cost: a Transformer-based model for high-precision semantic similarity and a Deep Averaging Network (DAN) for high-speed inference. Unlike previous methods trained on a single unsupervised task, USE is trained on a suite of multi-task learning objectives (including skip-thought-like, conversational response prediction, and natural language inference), ensuring the embeddings generalize well across a wide range of transfer tasks. This marked a shift toward "universal" pre-trained embeddings that could be plugged into any NLP pipeline.

\textbf{Sentence-BERT}. While BERT produces powerful contextual word embeddings, its native sentence representation (the [CLS] token) often yields poor performance for semantic similarity tasks, sometimes lagging behind simple GloVe averages. Reimers and Gurevych (2019) \cite{reimers2019sentence} addressed this with Sentence-BERT (S-BERT), which uses a siamese network structure to derive semantically meaningful sentence embeddings that can be compared using cosine similarity. By fine-tuning BERT on NLI (Natural Language Inference) and STS (Semantic Textual Similarity) data, S-BERT forces the vector space to align with human judgments of similarity. S-BERT reduced the time required to search for similar sentences from hours (with raw BERT) to milliseconds, enabling the massive adoption of dense retrieval systems and changing how semantic search is implemented in practice.

\subsection{Before vs.\ After GPT-3: Era Comparison}

The release of GPT-3 \cite{brown2020language} in mid-2020 marked a pivotal inflection point for the word embedding landscape. With 175 billion parameters and emergent few-shot learning capabilities, GPT-3 demonstrated that embeddings need not be standalone components but could function as internal representations within massive end-to-end systems. We therefore use GPT-3's release as the dividing line for a formal era comparison, partitioning our 149-paper dataset into a Pre-GPT-3 cohort ($\leq$ 2019, $n = 95$) and a Post-GPT-3 cohort ($\geq$ 2020, $n = 54$). Across seven hypothesis tests covering publication characteristics, collaboration patterns, and methodological focus, three yielded statistically significant differences while several others revealed suggestive trends. Quantitatively, the annual publication rate dropped from approximately 11.0 papers per year in the pre-GPT-3 period ($\leq$ 2019) to 9.0 papers per year in the post-GPT-3 period ($\geq$ 2020), an 18.2\% decrease.

\begin{figure}[h!]
    \centering
    \begin{tikzpicture}
        \begin{axis}[
            ybar,
            bar width=14pt,
            width=14cm,
            height=7cm,
            ymin=0,
            ymax=48,
            ylabel={Number of Papers},
            symbolic x coords={
                {Static Word Embeddings},
                {Contextual Word Embeddings},
                {Sentence/Document Embedding},
                {Statistical Representations}
            },
            xtick=data,
            xticklabel style={rotate=25, anchor=east, font=\footnotesize},
            nodes near coords,
            nodes near coords style={font=\footnotesize},
            enlarge x limits=0.15,
            area legend,
            legend style={at={(0.98,0.98)}, anchor=north east, font=\footnotesize},
            title={Embedding Categories: Pre vs.\ Post GPT-3}
        ]
        \addplot[fill=light-teal, draw=black] coordinates {
            ({Static Word Embeddings}, 40)
            ({Contextual Word Embeddings}, 19)
            ({Sentence/Document Embedding}, 17)
            ({Statistical Representations}, 19)
        };
        \addplot[fill=light-coral, draw=black] coordinates {
            ({Static Word Embeddings}, 10)
            ({Contextual Word Embeddings}, 26)
            ({Sentence/Document Embedding}, 17)
            ({Statistical Representations}, 1)
        };
        \legend{Pre-GPT-3 ($n = 95$), Post-GPT-3 ($n = 54$)}
        \end{axis}
    \end{tikzpicture}
    \caption{Distribution of embedding method categories before and after GPT-3. The post-GPT-3 era shows a dramatic shift away from static and statistical methods toward contextual and sentence/document embeddings.}
    \label{fig:gpt3_category_distribution}
\end{figure}
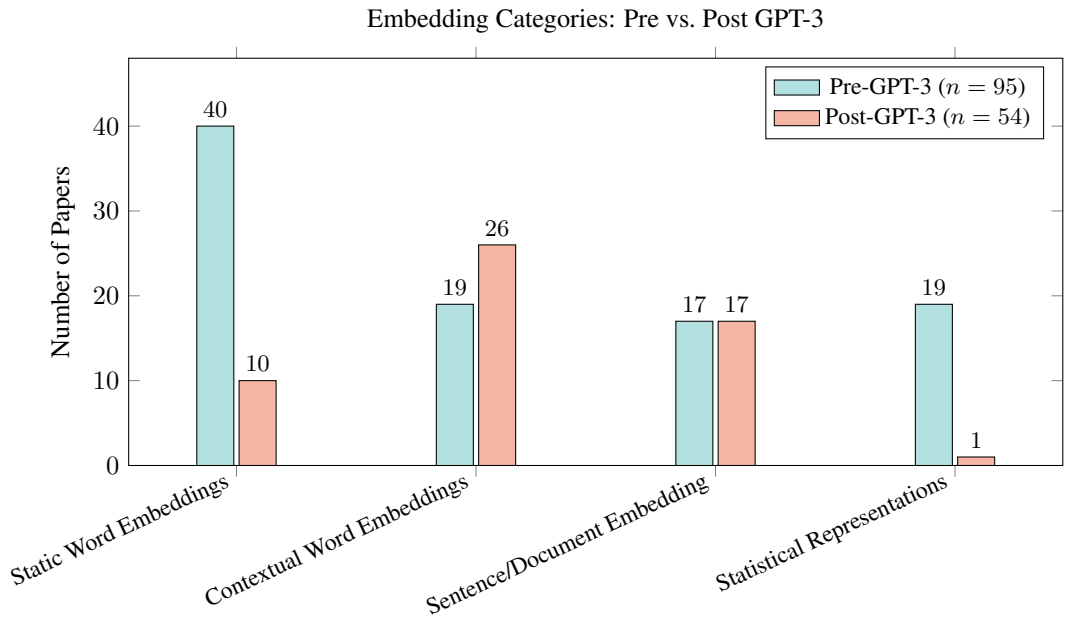

The most striking result concerns the shift in embedding category focus. A chi-square test of independence on the four-category distribution yields $\chi^{2}(3) = 25.97$, $p < 0.0001$, confirmed by Fisher's exact test ($p < 0.00001$), with a Cram\'{e}r's $V = 0.418$ indicating a medium effect. As shown in Figure~\ref{fig:gpt3_category_distribution}, static embeddings declined from 42.1\% of pre-GPT-3 papers to 18.5\% post-GPT-3, while statistical representations nearly vanished (20.0\% to 1.9\%). Conversely, contextual embeddings surged from 20.0\% to 48.1\%, and sentence/document embeddings rose from 17.9\% to 31.5\%. The odds ratio for contextual and sentence/document methods versus static and statistical methods is $6.4\times$ in the post-GPT-3 era, quantifying the magnitude of this paradigm shift.

Team size also increased significantly across eras. The mean number of authors per paper rose from 3.85 (pre-GPT-3) to 4.81 (post-GPT-3), a difference confirmed by Welch's $t$-test ($p = 0.018$) with Cohen's $d = 0.427$, a small-to-medium effect size. This finding complements the three-era descriptive analysis presented in Figure~\ref{fig:collaboration_dynamics} and suggests that the computational demands and interdisciplinary nature of post-GPT-3 research require larger collaborative teams.

Abstract length provides another significant marker of era-level change. Pre-GPT-3 abstracts average 138 words compared to 163 words in the post-GPT-3 period, a difference significant at $p = 0.0001$ (Welch's $t$-test) with Cohen's $d = 0.697$, a medium effect size. This increase likely reflects the growing methodological complexity of post-GPT-3 research, which often requires describing multiple pretraining objectives, fine-tuning strategies, and evaluation benchmarks within the abstract itself.

Several additional comparisons revealed suggestive but non-significant trends. Industry involvement rose from 35.8\% to 50.0\% (Fisher's exact, $p = 0.119$), a suggestive trend and consistent with the three-era pattern reported above. The share of original research papers declined from 65.3\% to 51.9\% (Fisher's exact, $p = 0.120$), suggesting a possible shift toward variant and enhancement studies in the post-GPT-3 period. International collaboration increased from 16.8\% to 25.9\% (Fisher's exact, $p = 0.206$), a trend that, while not statistically conclusive, aligns with the broader globalization of NLP research.

\noindent
The analysis of specific embedding techniques further quantifies the degree of methodological turnover across eras. Our dataset contains 64 unique techniques mentioned in pre-GPT-3 papers and 40 in post-GPT-3 papers, yet only 10 techniques appear in both periods. This remarkably low overlap indicates that the post-GPT-3 era introduced a largely new research vocabulary: 30 entirely novel techniques emerged, including SimCSE \cite{gao2021simcse}, LLM2Vec \cite{behnamghader2024llm2vec}, BigBird \cite{zaheer2020big}, E5 \cite{wang2022text}, T5 \cite{raffel2020exploring}, and CANINE \cite{clark2022canine}, while 54 pre-GPT-3 techniques received no further attention after 2019. This finding provides concrete evidence that GPT-3's release did not merely shift emphasis within the existing methodological toolkit but catalyzed a wholesale replacement of the techniques under investigation.

\noindent
Geographic concentration also shifted across eras. U.S.-affiliated papers declined from 63.2\% of the pre-GPT-3 corpus to 50.0\% in the post-GPT-3 period, and the number of unique contributing countries decreased from 19 to 15. While the U.S.\ remains the dominant contributor, this contraction, combined with the emergence of new affiliations from East Asian technology firms (Tencent, ByteDance, Meituan) and universities (Hong Kong Polytechnic University, Korea University), suggests a partial redistribution of research activity toward industry labs in the Asia-Pacific region, even as total geographic diversity slightly narrowed.

\subsubsection{Interaction Analyses}

Beyond the marginal comparisons above, we examine three cross-tabulations that reveal how the GPT-3 transition played out differently across embedding categories.

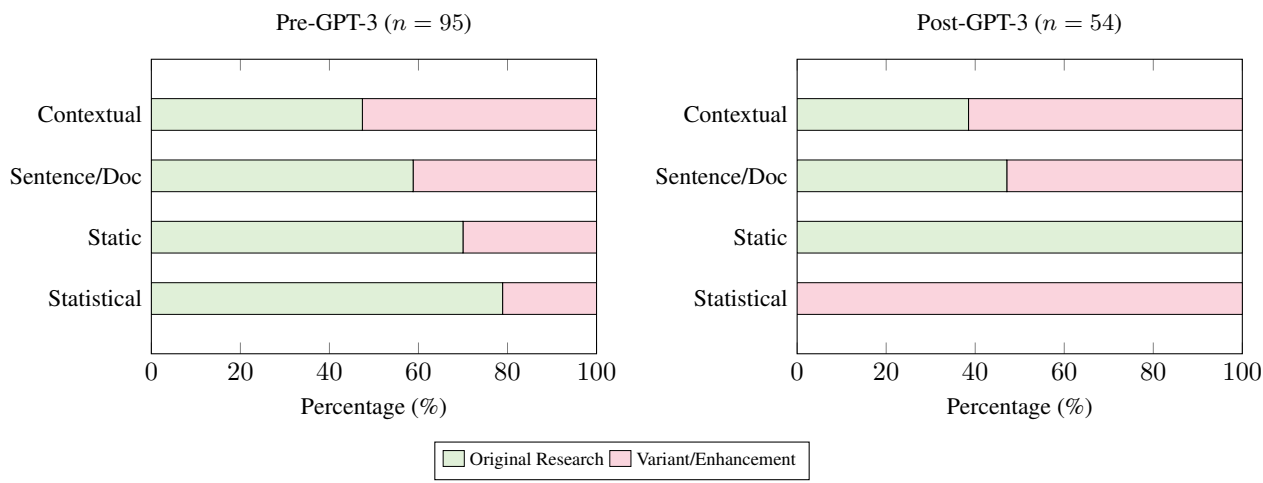
\begin{figure}[h!]
    \centering
    \begin{subfigure}[t]{0.48\textwidth}
        \centering
        \begin{tikzpicture}
            \begin{axis}[
                xbar stacked,
                bar width=12pt,
                width=7.5cm,
                height=5.5cm,
                xmin=0, xmax=100,
                xlabel={\small Percentage (\%)},
                title={\footnotesize Pre-GPT-3 ($n = 95$)},
                symbolic y coords={Statistical, Static, {Sentence/Doc}, Contextual},
                ytick=data,
                yticklabel style={font=\footnotesize},
                enlarge y limits=0.3,
                legend to name=grouplegend14,
                legend columns=2,
                legend style={font=\scriptsize},
            ]
            \addplot[fill=light-green, draw=black, line width=0.3pt] coordinates
                {(78.9,Statistical) (70.0,Static) (58.8,{Sentence/Doc}) (47.4,Contextual)};
            \addplot[fill=light-pink, draw=black, line width=0.3pt] coordinates
                {(21.1,Statistical) (30.0,Static) (41.2,{Sentence/Doc}) (52.6,Contextual)};
            \node[font=\scriptsize\bfseries] at (axis cs:39.45,Statistical) {79\%};
            \node[font=\scriptsize\bfseries] at (axis cs:35.0,Static) {70\%};
            \node[font=\scriptsize\bfseries] at (axis cs:29.4,{Sentence/Doc}) {59\%};
            \node[font=\scriptsize\bfseries] at (axis cs:23.7,Contextual) {47\%};
            \node[font=\scriptsize\bfseries] at (axis cs:89.45,Statistical) {21\%};
            \node[font=\scriptsize\bfseries] at (axis cs:85.0,Static) {30\%};
            \node[font=\scriptsize\bfseries] at (axis cs:79.4,{Sentence/Doc}) {41\%};
            \node[font=\scriptsize\bfseries] at (axis cs:73.7,Contextual) {53\%};
            \legend{Original Research, Variant/Enhancement}
            \end{axis}
        \end{tikzpicture}
    \end{subfigure}
    \hfill
    \begin{subfigure}[t]{0.48\textwidth}
        \centering
        \begin{tikzpicture}
            \begin{axis}[
                xbar stacked,
                bar width=12pt,
                width=7.5cm,
                height=5.5cm,
                xmin=0, xmax=100,
                xlabel={\small Percentage (\%)},
                title={\footnotesize Post-GPT-3 ($n = 54$)},
                symbolic y coords={Statistical, Static, {Sentence/Doc}, Contextual},
                ytick=data,
                yticklabel style={font=\footnotesize},
                enlarge y limits=0.3,
            ]
            \addplot[fill=light-green, draw=black, line width=0.3pt] coordinates
                {(0,Statistical) (100,Static) (47.1,{Sentence/Doc}) (38.5,Contextual)};
            \addplot[fill=light-pink, draw=black, line width=0.3pt] coordinates
                {(100,Statistical) (0,Static) (52.9,{Sentence/Doc}) (61.5,Contextual)};
            \node[font=\scriptsize\bfseries] at (axis cs:50.0,Statistical) {100\%};
            \node[font=\scriptsize\bfseries] at (axis cs:50.0,Static) {100\%};
            \node[font=\scriptsize\bfseries] at (axis cs:23.55,{Sentence/Doc}) {47\%};
            \node[font=\scriptsize\bfseries] at (axis cs:73.55,{Sentence/Doc}) {53\%};
            \node[font=\scriptsize\bfseries] at (axis cs:19.25,Contextual) {38\%};
            \node[font=\scriptsize\bfseries] at (axis cs:69.25,Contextual) {62\%};
            \end{axis}
        \end{tikzpicture}
    \end{subfigure}

    \vspace{4pt}
    \centering\pgfplotslegendfromname{grouplegend14}
    \caption{Original vs.\ Variant/Enhancement by embedding category: Pre- vs.\ Post-GPT-3. Post-GPT-3 contextual embeddings are dominated by variant/enhancement work (62\%), while the few remaining static embedding papers are entirely original contributions.}
    \label{fig:cat_originality}
\end{figure}

\paragraph{Category $\times$ Originality.}
Figure~\ref{fig:cat_originality} cross-tabulates embedding category with research type (original vs.\ variant/enhancement) for each era. In the pre-GPT-3 period, static embeddings were predominantly original contributions (28 original vs.\ 12 variants), and statistical methods likewise favored original work (15 vs.\ 4). After GPT-3, contextual embeddings became the dominant category but shifted toward enhancement work: 10 original papers versus 16 variants, yielding an originality rate of only 38.5\% compared to 47.4\% in the pre-GPT-3 contextual cohort. Meanwhile, all 10 post-GPT-3 static embedding papers were classified as original research with zero variants suggesting that researchers who continued working on static methods post-GPT-3 did so only when proposing genuinely novel approaches. This pattern indicates that the overall decline in originality (65.3\% to 51.9\%) is primarily concentrated in the contextual embedding category, where the maturation of transformer-based architectures has generated a large body of incremental refinement work.

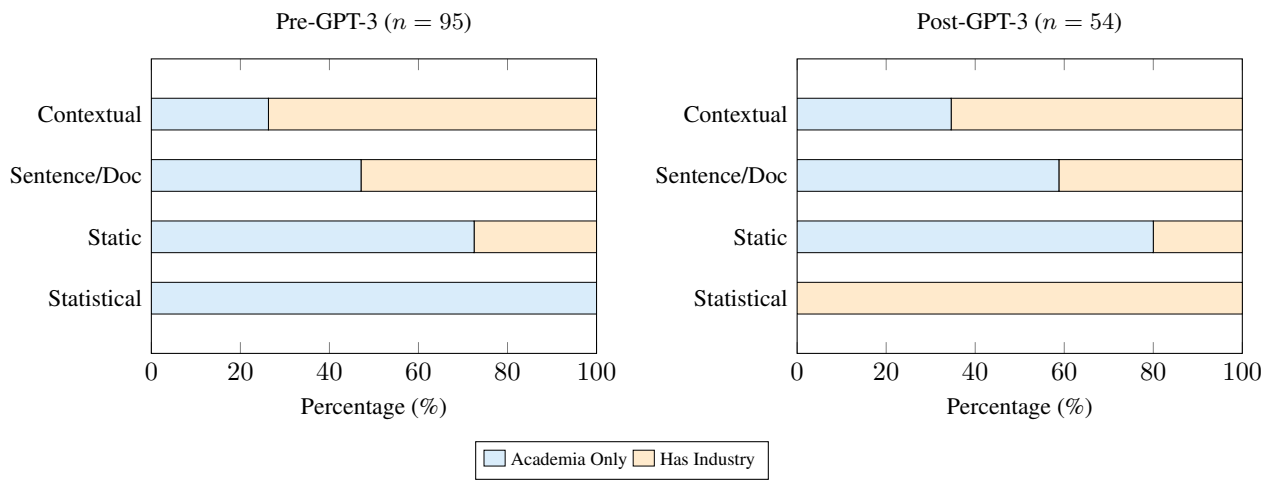
\begin{figure}[h!]
    \centering
    \begin{subfigure}[t]{0.48\textwidth}
        \centering
        \begin{tikzpicture}
            \begin{axis}[
                xbar stacked,
                bar width=12pt,
                width=7.5cm,
                height=5.5cm,
                xmin=0, xmax=100,
                xlabel={\small Percentage (\%)},
                title={\footnotesize Pre-GPT-3 ($n = 95$)},
                symbolic y coords={Statistical, Static, {Sentence/Doc}, Contextual},
                ytick=data,
                yticklabel style={font=\footnotesize},
                enlarge y limits=0.3,
                legend to name=grouplegend15,
                legend columns=2,
                legend style={font=\scriptsize},
            ]
            \addplot[fill=light-blue, draw=black, line width=0.3pt] coordinates
                {(100,Statistical) (72.5,Static) (47.1,{Sentence/Doc}) (26.3,Contextual)};
            \addplot[fill=light-orange, draw=black, line width=0.3pt] coordinates
                {(0,Statistical) (27.5,Static) (52.9,{Sentence/Doc}) (73.7,Contextual)};
            \node[font=\scriptsize\bfseries] at (axis cs:50.0,Statistical) {100\%};
            \node[font=\scriptsize\bfseries] at (axis cs:36.25,Static) {72\%};
            \node[font=\scriptsize\bfseries] at (axis cs:23.55,{Sentence/Doc}) {47\%};
            \node[font=\scriptsize\bfseries] at (axis cs:13.15,Contextual) {26\%};
            \node[font=\scriptsize\bfseries] at (axis cs:86.25,Static) {28\%};
            \node[font=\scriptsize\bfseries] at (axis cs:73.55,{Sentence/Doc}) {53\%};
            \node[font=\scriptsize\bfseries] at (axis cs:63.15,Contextual) {74\%};
            \legend{Academia Only, Has Industry}
            \end{axis}
        \end{tikzpicture}
    \end{subfigure}
    \hfill
    \begin{subfigure}[t]{0.48\textwidth}
        \centering
        \begin{tikzpicture}
            \begin{axis}[
                xbar stacked,
                bar width=12pt,
                width=7.5cm,
                height=5.5cm,
                xmin=0, xmax=100,
                xlabel={\small Percentage (\%)},
                title={\footnotesize Post-GPT-3 ($n = 54$)},
                symbolic y coords={Statistical, Static, {Sentence/Doc}, Contextual},
                ytick=data,
                yticklabel style={font=\footnotesize},
                enlarge y limits=0.3,
            ]
            \addplot[fill=light-blue, draw=black, line width=0.3pt] coordinates
                {(0,Statistical) (80.0,Static) (58.8,{Sentence/Doc}) (34.6,Contextual)};
            \addplot[fill=light-orange, draw=black, line width=0.3pt] coordinates
                {(100,Statistical) (20.0,Static) (41.2,{Sentence/Doc}) (65.4,Contextual)};
            \node[font=\scriptsize\bfseries] at (axis cs:50.0,Statistical) {100\%};
            \node[font=\scriptsize\bfseries] at (axis cs:40.0,Static) {80\%};
            \node[font=\scriptsize\bfseries] at (axis cs:90.0,Static) {20\%};
            \node[font=\scriptsize\bfseries] at (axis cs:29.4,{Sentence/Doc}) {59\%};
            \node[font=\scriptsize\bfseries] at (axis cs:79.4,{Sentence/Doc}) {41\%};
            \node[font=\scriptsize\bfseries] at (axis cs:17.3,Contextual) {35\%};
            \node[font=\scriptsize\bfseries] at (axis cs:67.3,Contextual) {65\%};
            \end{axis}
        \end{tikzpicture}
    \end{subfigure}

    \vspace{4pt}
    \centering\pgfplotslegendfromname{grouplegend15}
    \caption{Industry involvement by embedding category: Pre- vs.\ Post-GPT-3. Contextual embeddings are heavily industry-driven in both eras, while static and statistical methods remain predominantly academic. The overall rise in industry involvement is largely a compositional effect of the field's shift toward contextual methods.}
    \label{fig:cat_industry}
\end{figure}

\paragraph{Category $\times$ Industry Involvement.}
Figure~\ref{fig:cat_industry} breaks down industry involvement by embedding category for each era. A striking asymmetry emerges: in both eras, contextual embedding research is heavily industry-driven (73.7\% pre-GPT-3, 65.4\% post-GPT-3), whereas static embeddings are predominantly academic (72.5\% academia-only pre-GPT-3, 80.0\% post-GPT-3). Statistical representations had zero industry involvement before GPT-3, consistent with their origins in academic linguistics and information retrieval. (The post-GPT-3 statistical category contains only $n = 1$ paper, so its 100\% industry rate shown in Figure~\ref{fig:cat_industry} is not interpretable as a trend.) Sentence/document embeddings occupy a middle ground, with industry involvement declining modestly from 52.9\% to 41.2\% across eras. These patterns reveal a Simpson's Paradox-like phenomenon: within each embedding category, industry involvement rates remained relatively stable or even slightly declined, yet the aggregate industry share rose from 35.8\% to 50.0\%. The explanation is compositional; the field abandoned the categories that were historically academic-friendly (static and statistical methods) and migrated toward resource-intensive architectures (contextual embeddings) where industry has always been the dominant contributor. In other words, the apparent ``industry takeover'' of word embedding research is driven not by industry labs displacing academics within existing research topics, but by the field's wholesale shift toward capital-intensive paradigms that inherently favor well-resourced industrial labs.

\begin{figure}[h!]
    \centering
    \begin{tikzpicture}
        \begin{axis}[
            ybar,
            bar width=15pt,
            width=14cm,
            height=7cm,
            ymin=0,
            ymax=7.5,
            ylabel={Mean Number of Authors},
            symbolic x coords={
                {Contextual Word Embeddings},
                {Sentence/Document Embedding},
                {Static Word Embeddings},
                {Statistical Representations}
            },
            xtick=data,
            xticklabel style={rotate=25, anchor=east, font=\footnotesize},
            nodes near coords,
            nodes near coords style={font=\footnotesize, /pgf/number format/fixed, /pgf/number format/precision=1},
            enlarge x limits=0.15,
            area legend,
            legend style={at={(0.5,0.98)}, anchor=north, legend columns=2, font=\footnotesize},
            title={Mean Team Size by Category and Era}
        ]
        \addplot[fill=light-teal, draw=black] coordinates {
            ({Contextual Word Embeddings}, 4.79)
            ({Sentence/Document Embedding}, 3.94)
            ({Static Word Embeddings}, 4.05)
            ({Statistical Representations}, 2.42)
        };
        \addplot[fill=light-coral, draw=black] coordinates {
            ({Contextual Word Embeddings}, 5.46)
            ({Sentence/Document Embedding}, 4.71)
            ({Static Word Embeddings}, 3.20)
            ({Statistical Representations}, 6.00)
        };
        \legend{Pre-GPT-3, Post-GPT-3}
        \end{axis}
    \end{tikzpicture}
    \caption{Mean team size by embedding category and era. Contextual embeddings consistently require the largest teams, while static embedding team sizes actually decreased post-GPT-3, indicating the overall team size increase is driven by the field's compositional shift toward resource-intensive categories.}
    \label{fig:teamsize_category}
\end{figure}
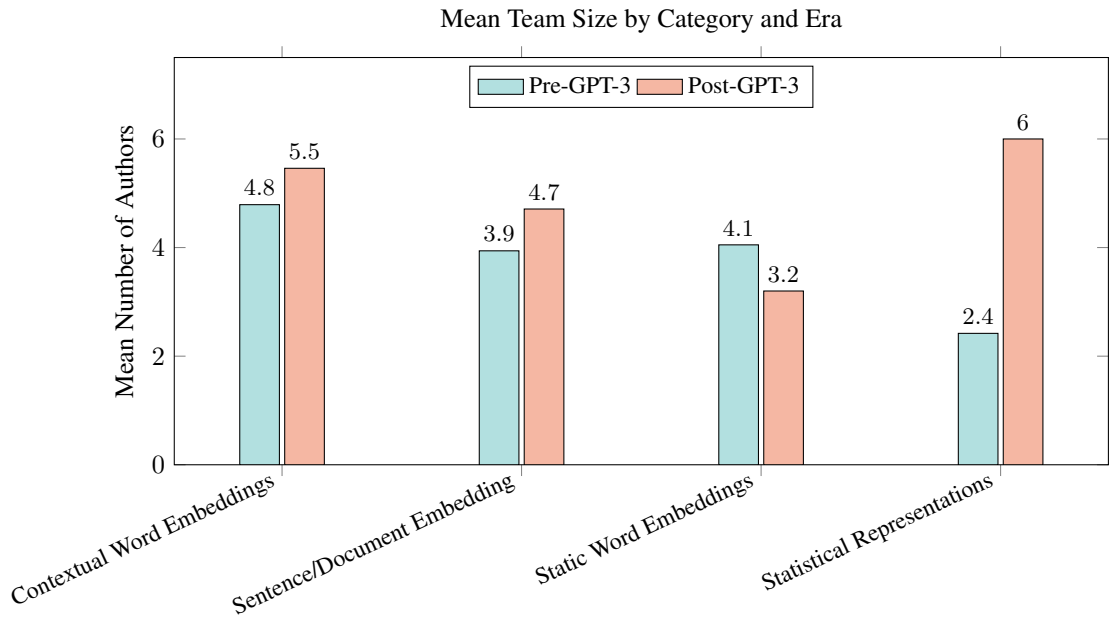

\paragraph{Team Size $\times$ Category.}
Figure~\ref{fig:teamsize_category} reports mean team size by embedding category and era. Contextual embedding papers consistently have the largest teams (4.79 pre-GPT-3, 5.46 post-GPT-3), reflecting the computational infrastructure required for transformer-based research. The only statistically significant within-category change is for static embeddings, where mean team size decreased from 4.05 to 3.20 (Welch's $t$-test, $p = 0.046$), which is the opposite direction of the overall trend. This suggests that the few researchers who continued pursuing static methods after GPT-3 were small, specialized teams, while the overall increase in team size is driven by the growing dominance of contextual and sentence-level embedding research, which requires larger collaborative efforts. Sentence/document embeddings also show a notable increase (3.94 to 4.71), though this does not reach significance ($p = 0.391$).

Table~\ref{tab:gpt3_hypothesis_tests} summarizes all seven hypothesis tests conducted in this era comparison analysis. We note that no formal correction for multiple comparisons (e.g., Bonferroni) was applied. With seven tests at $\alpha = 0.05$, the family-wise error rate is approximately 30\%, so the results, particularly those with $p$-values near the conventional threshold, should be interpreted as exploratory rather than confirmatory. The three tests yielding $p < 0.02$ (category distribution, team size, abstract length) would survive a Bonferroni-corrected threshold of $\alpha/7 \approx 0.0071$ only in the case of category distribution ($p < 0.0001$) and abstract length ($p = 0.0001$); the team size result ($p = 0.018$) would not, though it remains suggestive.

\begin{table}[h!]
\centering
\small
\caption{Summary of hypothesis tests comparing Pre-GPT-3 ($\leq$ 2019, $n = 95$) vs.\ Post-GPT-3 ($\geq$ 2020, $n = 54$) papers.}
\label{tab:gpt3_hypothesis_tests}
\begin{tabular}{@{}l l l l l l@{}}
\toprule
\textbf{Metric} & \textbf{Pre-GPT-3} & \textbf{Post-GPT-3} & \textbf{Test} & \textbf{$p$-value} & \textbf{Sig.} \\
\midrule
Category distribution & -- & -- & Chi-square / Fisher's & $<$ 0.0001 & Yes \\
Team size (mean) & 3.85 & 4.81 & Welch's $t$ & 0.018 & Yes \\
Abstract length (words) & 138 & 163 & Welch's $t$ & 0.0001 & Yes \\
Industry involvement (\%) & 35.8 & 50.0 & Fisher's exact & 0.119 & No \\
Original research (\%) & 65.3 & 51.9 & Fisher's exact & 0.120 & No \\
International collab.\ (\%) & 16.8 & 25.9 & Fisher's exact & 0.206 & No \\
Publication venue dist. & -- & -- & Fisher's exact & 0.855 & No \\
\bottomrule
\end{tabular}
\end{table}

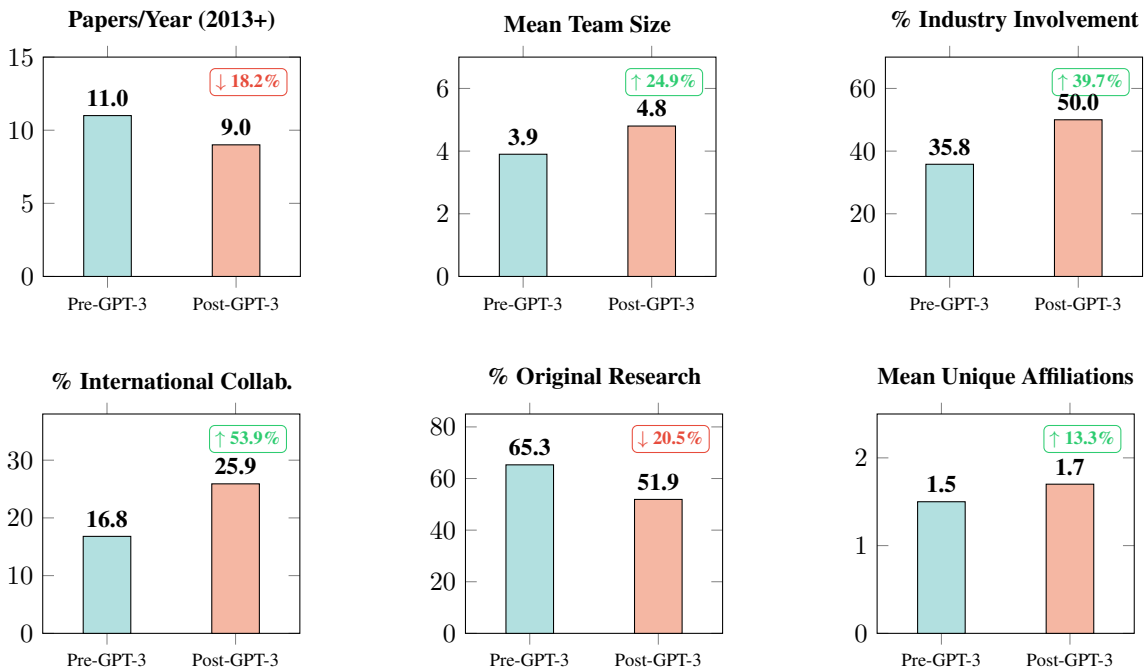
\begin{figure}[h!]
    \centering
    \begin{subfigure}[t]{0.32\textwidth}
        \centering
        \begin{tikzpicture}
            \begin{axis}[
                ybar, bar width=18pt, width=5cm, height=4.5cm,
                ymin=0, ymax=15,
                symbolic x coords={Pre-GPT-3, Post-GPT-3},
                xtick={Pre-GPT-3, Post-GPT-3},
                xticklabel style={font=\scriptsize}, enlarge x limits=0.5,
                title={\footnotesize\bfseries Papers/Year (2013+)},
            ]
            \addplot[fill=light-teal, draw=black, bar shift=0pt,
                nodes near coords, point meta=explicit symbolic,
                nodes near coords style={font=\footnotesize\bfseries}]
                coordinates {(Pre-GPT-3, 11.0) [11.0]};
            \addplot[fill=light-coral, draw=black, bar shift=0pt,
                nodes near coords, point meta=explicit symbolic,
                nodes near coords style={font=\footnotesize\bfseries}]
                coordinates {(Post-GPT-3, 9.0) [9.0]};
            \node[font=\scriptsize\bfseries, text=variant-red, fill=white, draw=variant-red,
                rounded corners=2pt, inner sep=2pt, anchor=north east]
                at (axis description cs:0.95,0.95) {$\downarrow$ 18.2\%};
            \end{axis}
        \end{tikzpicture}
    \end{subfigure}
    \hfill
    \begin{subfigure}[t]{0.32\textwidth}
        \centering
        \begin{tikzpicture}
            \begin{axis}[
                ybar, bar width=18pt, width=5cm, height=4.5cm,
                ymin=0, ymax=7,
                symbolic x coords={Pre-GPT-3, Post-GPT-3},
                xtick={Pre-GPT-3, Post-GPT-3},
                xticklabel style={font=\scriptsize}, enlarge x limits=0.5,
                title={\footnotesize\bfseries Mean Team Size},
            ]
            \addplot[fill=light-teal, draw=black, bar shift=0pt,
                nodes near coords, point meta=explicit symbolic,
                nodes near coords style={font=\footnotesize\bfseries}]
                coordinates {(Pre-GPT-3, 3.9) [3.9]};
            \addplot[fill=light-coral, draw=black, bar shift=0pt,
                nodes near coords, point meta=explicit symbolic,
                nodes near coords style={font=\footnotesize\bfseries}]
                coordinates {(Post-GPT-3, 4.8) [4.8]};
            \node[font=\scriptsize\bfseries, text=orig-green, fill=white, draw=orig-green,
                rounded corners=2pt, inner sep=2pt, anchor=north east]
                at (axis description cs:0.95,0.95) {$\uparrow$ 24.9\%};
            \end{axis}
        \end{tikzpicture}
    \end{subfigure}
    \hfill
    \begin{subfigure}[t]{0.32\textwidth}
        \centering
        \begin{tikzpicture}
            \begin{axis}[
                ybar, bar width=18pt, width=5cm, height=4.5cm,
                ymin=0, ymax=70,
                symbolic x coords={Pre-GPT-3, Post-GPT-3},
                xtick={Pre-GPT-3, Post-GPT-3},
                xticklabel style={font=\scriptsize}, enlarge x limits=0.5,
                title={\footnotesize\bfseries \% Industry Involvement},
            ]
            \addplot[fill=light-teal, draw=black, bar shift=0pt,
                nodes near coords, point meta=explicit symbolic,
                nodes near coords style={font=\footnotesize\bfseries}]
                coordinates {(Pre-GPT-3, 35.8) [35.8]};
            \addplot[fill=light-coral, draw=black, bar shift=0pt,
                nodes near coords, point meta=explicit symbolic,
                nodes near coords style={font=\footnotesize\bfseries}]
                coordinates {(Post-GPT-3, 50.0) [50.0]};
            \node[font=\scriptsize\bfseries, text=orig-green, fill=white, draw=orig-green,
                rounded corners=2pt, inner sep=2pt, anchor=north east]
                at (axis description cs:0.95,0.95) {$\uparrow$ 39.7\%};
            \end{axis}
        \end{tikzpicture}
    \end{subfigure}

    \vspace{0.5cm}

    \begin{subfigure}[t]{0.32\textwidth}
        \centering
        \begin{tikzpicture}
            \begin{axis}[
                ybar, bar width=18pt, width=5cm, height=4.5cm,
                ymin=0, ymax=38,
                symbolic x coords={Pre-GPT-3, Post-GPT-3},
                xtick={Pre-GPT-3, Post-GPT-3},
                xticklabel style={font=\scriptsize}, enlarge x limits=0.5,
                title={\footnotesize\bfseries \% International Collab.},
            ]
            \addplot[fill=light-teal, draw=black, bar shift=0pt,
                nodes near coords, point meta=explicit symbolic,
                nodes near coords style={font=\footnotesize\bfseries}]
                coordinates {(Pre-GPT-3, 16.8) [16.8]};
            \addplot[fill=light-coral, draw=black, bar shift=0pt,
                nodes near coords, point meta=explicit symbolic,
                nodes near coords style={font=\footnotesize\bfseries}]
                coordinates {(Post-GPT-3, 25.9) [25.9]};
            \node[font=\scriptsize\bfseries, text=orig-green, fill=white, draw=orig-green,
                rounded corners=2pt, inner sep=2pt, anchor=north east]
                at (axis description cs:0.95,0.95) {$\uparrow$ 53.9\%};
            \end{axis}
        \end{tikzpicture}
    \end{subfigure}
    \hfill
    \begin{subfigure}[t]{0.32\textwidth}
        \centering
        \begin{tikzpicture}
            \begin{axis}[
                ybar, bar width=18pt, width=5cm, height=4.5cm,
                ymin=0, ymax=85,
                symbolic x coords={Pre-GPT-3, Post-GPT-3},
                xtick={Pre-GPT-3, Post-GPT-3},
                xticklabel style={font=\scriptsize}, enlarge x limits=0.5,
                title={\footnotesize\bfseries \% Original Research},
            ]
            \addplot[fill=light-teal, draw=black, bar shift=0pt,
                nodes near coords, point meta=explicit symbolic,
                nodes near coords style={font=\footnotesize\bfseries}]
                coordinates {(Pre-GPT-3, 65.3) [65.3]};
            \addplot[fill=light-coral, draw=black, bar shift=0pt,
                nodes near coords, point meta=explicit symbolic,
                nodes near coords style={font=\footnotesize\bfseries}]
                coordinates {(Post-GPT-3, 51.9) [51.9]};
            \node[font=\scriptsize\bfseries, text=variant-red, fill=white, draw=variant-red,
                rounded corners=2pt, inner sep=2pt, anchor=north east]
                at (axis description cs:0.95,0.95) {$\downarrow$ 20.5\%};
            \end{axis}
        \end{tikzpicture}
    \end{subfigure}
    \hfill
    \begin{subfigure}[t]{0.32\textwidth}
        \centering
        \begin{tikzpicture}
            \begin{axis}[
                ybar, bar width=18pt, width=5cm, height=4.5cm,
                ymin=0, ymax=2.5,
                symbolic x coords={Pre-GPT-3, Post-GPT-3},
                xtick={Pre-GPT-3, Post-GPT-3},
                xticklabel style={font=\scriptsize}, enlarge x limits=0.5,
                title={\footnotesize\bfseries Mean Unique Affiliations},
            ]
            \addplot[fill=light-teal, draw=black, bar shift=0pt,
                nodes near coords, point meta=explicit symbolic,
                nodes near coords style={font=\footnotesize\bfseries}]
                coordinates {(Pre-GPT-3, 1.5) [1.5]};
            \addplot[fill=light-coral, draw=black, bar shift=0pt,
                nodes near coords, point meta=explicit symbolic,
                nodes near coords style={font=\footnotesize\bfseries}]
                coordinates {(Post-GPT-3, 1.7) [1.7]};
            \node[font=\scriptsize\bfseries, text=orig-green, fill=white, draw=orig-green,
                rounded corners=2pt, inner sep=2pt, anchor=north east]
                at (axis description cs:0.95,0.95) {$\uparrow$ 13.3\%};
            \end{axis}
        \end{tikzpicture}
    \end{subfigure}
    \caption{Summary dashboard comparing six key metrics across the Pre-GPT-3 and Post-GPT-3 eras. Each panel shows the pre- and post-GPT-3 values with the percentage change annotated. Green arrows ($\uparrow$) indicate increases; red arrows ($\downarrow$) indicate decreases.}
    \label{fig:gpt3_summary_dashboard}
\end{figure}

\section{Discussion}

The trends identified in our analysis highlight several critical shifts in the NLP landscape, with implications for research practice, global equity, and ethical AI development.

\subsection{Implications for Researchers: Navigating the LLM Paradigm}

The transition from standalone word embeddings to internal representations within Large Language Models (LLMs) fundamentally redefines the research agenda for NLP practitioners. As our analysis shows, mean team sizes have grown from 2.8 authors in the pre-Word2Vec era to 4.8 in the post-GPT-3 period (Figure~\ref{fig:collaboration_dynamics}), and industry involvement has risen from 6\% to 50\% across the same timeframe. The near-complete methodological turnover documented in Section~4, where only 10 of 94 unique techniques span both eras, confirms that the field has moved from small-scale academic teams to large, industry-led collaborations, largely due to the immense computational resources required for modern contextual models. For researchers, this shift implies a need to move beyond merely optimizing vector dimensions or local context windows. Instead, future efforts should prioritize interpretability and efficiency. Understanding how these internal embeddings encode reasoning patterns or discourse structure remains a critical frontier. Moreover, the historical continuity identified, from TF-IDF’s informativeness to BERT’s bidirectional context, suggests that foundational linguistic principles still underpin modern architectures. Researchers should continue to look back at these primitive methods to demystify the black box of LLMs and develop lighter, more sustainable models for low-resource environments. A concrete illustration of this principle is the recent renaissance of statistical retrieval methods in Retrieval-Augmented Generation (RAG) pipelines: sparse methods such as TF-IDF and BM25 are now widely used as efficient first-stage retrievers alongside dense vector search, because their exact keyword matching complements the semantic generalization of neural embeddings. This resurgence demonstrates that ``primitive'' representations retain practical value even in the LLM era, and that hybrid architectures combining classical and neural components often outperform purely neural systems in retrieval-intensive tasks.

\subsection{Societal Impact: Resource Disparity and Innovation Biases}

The evolution of word embedding techniques carries significant societal implications regarding global technological equity. Our data highlights a stark national imbalance: the United States alone accounts for 378 out of 614 total author-affiliations in the dataset (Figure~\ref{fig:distribution_nations}), and just five countries (U.S., China, UK, Germany, Canada) represent over 80\% of all contributions. At the institutional level, a single company like Google appears in 95 author-affiliations, more than double the next contributor (Figure~\ref{fig:distribution_affiliations}). This heavy concentration, fueled by investments from major technology corporations, has significant implications: it can lead to innovation biases, where embedding models are optimized for high-resource languages and specific cultural contexts, potentially marginalizing minority languages and diverse perspectives. As embeddings become the foundation of modern information retrieval, sentiment analysis, and generative AI, the barriers to entry, such as limited computational infrastructure in other nations, could widen the global digital divide. Addressing these disparities requires a shift toward more inclusive, international collaborations that prioritize multilingual embeddings and open-source accessibility.

\subsection{Ethical Considerations in Semantic Representation}

Finally, the shift toward deep contextual and generative models like GPT and BERT, which now account for 48.1\% of post-GPT-3 publications in our dataset (Figure~\ref{fig:gpt3_category_distribution}), introduces complex ethical challenges regarding the nature of truth and representation. While early static embeddings like Word2Vec were criticized for inheriting human biases (e.g., gendered analogies), modern contextual embeddings are even more opaque. These models do not just represent words; they model entire probability distributions of human thought, which can inadvertently amplify systemic biases or generate hallucinated information. The societal impact is profound: as these embeddings power the tools we use for education, law, and communication, the lack of an explicit bidirectional context in some models or the sensitivity to prompt design in others can lead to inconsistent or harmful outputs. It is incumbent upon the research community to develop robust evaluative frameworks that ensure semantic representations remain aligned with ethical standards and human values.

\section{Conclusion}

This study has traced the seven-decade evolution of word embedding techniques, from Harris's foundational work on distributional structure in 1954 to the transformer-based contextual models of the 2020s. By assembling and analyzing a curated dataset of 149 papers, we have provided both a comprehensive methodological survey and a data-driven bibliometric analysis that, to our knowledge, is the first to combine these two perspectives for the word embedding literature. The historical trajectory reveals a clear progression: sparse, high-dimensional statistical representations (Bag-of-Words, TF-IDF, LSA) gave way to dense, fixed-dimensional static vectors (Word2Vec, GloVe, FastText), which in turn were superseded by dynamic, context-dependent models (ELMo, BERT, GPT). Each generation addressed the specific limitations of its predecessors, dense vectors solved the sparsity and lack of similarity structure in one-hot encoding; subword information solved the out-of-vocabulary problem; and attention mechanisms solved the polysemy problem that plagued static models.

Beyond methodological trends, our bibliometric analysis documents the increasing role of industry in word embedding research. Industry involvement rose from 6\% in the earliest period (1954 -- 2012) to 50\% in the LLM era (2020--2025), and contextual embedding research is heavily industry-driven in both the pre- and post-GPT-3 periods. Geographic analysis reveals that the United States accounts for over half of all author affiliations, with research activity concentrated in a small number of resource-rich nations. These patterns underscore the growing resource barriers to entry in embedding research and the need for more inclusive international collaboration, particularly for multilingual and low-resource language settings.

Several limitations of this work should be acknowledged. First, all annotation was performed by a single researcher without formal inter-rater reliability assessment, which may introduce classification bias despite consistent application of the defined criteria. Second, the dataset of 149 papers, while carefully curated to include methods that introduced or significantly extended embedding techniques, does not capture the full universe of applied studies that use embeddings in downstream tasks. Third, our restriction to English-language publications reflects the dominance of English in NLP venues but may underrepresent contributions from non-English-speaking research communities. Fourth, the binary pre/post GPT-3 division, while analytically useful, necessarily simplifies what was a gradual transition rather than an abrupt shift.

Looking ahead, several directions for future research emerge from our analysis. First, the near-total replacement of pre-GPT-3 techniques raises the question of whether valuable ideas from earlier paradigms, such as explicit co-occurrence modeling, hyperbolic geometry, or syntactic dependency structures, might be productively revisited within modern architectures. Second, the growing dominance of industry-led research calls for continued monitoring of how commercial interests shape the methodological agenda of the field. Third, as embeddings become increasingly internalized within large language models, developing methods to interpret, probe, and evaluate these internal representations remains a critical frontier. Finally, the concentration of research in a small number of countries and institutions suggests that expanding access to computational resources and fostering diverse, international collaboration will be essential for ensuring that embedding technologies serve a broad range of languages, cultures, and communities.

\section*{Data and Code Availability}
The annotated dataset of 149 word embedding papers and all analysis code used in this study are publicly available at \url{https://github.com/minhanhnguy/tracing-the-evolution-of-word-embeddings-in-nlp}. The repository includes the complete dataset with author affiliations, embedding categories, and all bibliometric annotations, as well as the Jupyter notebook and visualization scripts used to reproduce the statistical analyses and figures presented in this paper.

\bibliographystyle{unsrt}
\bibliography{references}  

\end{document}